\documentclass[usenatbib,onecolumn]{mn2essr}
\usepackage{graphicx}
\graphicspath{{./images/}}
\usepackage{tabularx}
\usepackage{amsmath}
\newcommand{\beq}{\begin{equation}}
\newcommand{\eeq}{\end{equation}}

\def\ssr{Sp Sci. Rev.}

\def\bar{\overline}

\def\pre{Phys. Rev E.}
\def\mnras{MNRAS}
\def\apj{ApJ}
\def\apjl{ApJL}
\def\aap{A\&A}

\def\jcap{Journal of Cosmology and Astroparticle Physics}

\def\emf{\overline{\mbox{${\cal E}$}} {}}
\def\emfb{\overline{\mbox{\boldmath ${\cal E}$}} {}}
\def\bbE{\bar {\bf E}}

\def\beq{\begin{equation}}
\def\ee{\end{equation}}
\def\lsim{\mathrel{\rlap{\lower4pt\hbox{\hskip1pt$\sim$}}
    \raise1pt\hbox{$<$}}}
\def\gsim{\mathrel{\rlap{\lower4pt\hbox{\hskip1pt$\sim$}}
    \raise1pt\hbox{$>$}}}
\def\bfE{{\bf E}}
\def\bfJ{{\bf J}}

\def\bfA{{\bf A}}
\def\bfa{{\bf a}}
\def\bfe{{\bf e}}
\def\bfB{{\bf B}}
\def\bbJ{\bar {\bf J}}

\def\bB{\overline B}
\def\ts{\times}
\def\lb{\langle}
\def\rb{\rangle}
\def\curl{\nabla {\ts}}

\def\bfv{{\bf v}}

\def\bfV{{\bf V}}
\def\bfj{{\bf j}}

\def\bfe{{\bf e}}

\def\bfb{{\bf b}}

\def\bfB{{\bf B}}
\def\bfA{{\bf A}}

\def\bbB{\overline {\bf B}}
\def\bbA{\overline {\bf A}}

\def\div{\nabla\cdot}

\def\la{\lambda}

\title[Magnetic Helicity and Large Scale Magnetic Fields: A Primer]
 {Magnetic Helicity and Large Scale Magnetic Fields: A Primer}
\author [Eric G. Blackman]{Eric G. Blackman$^{1}$\thanks{E-mail: blackman@pas.rochester.edu} \\ $^{1}$Department of Physics and Astronomy, University of Rochester, Rochester NY, 14618, USA\\}
\begin{document}
\date{}
\maketitle
\label{firstpage}

\begin{abstract}
Magnetic fields of   laboratory, planetary, stellar, and galactic plasmas commonly  exhibit significant order on  large temporal or spatial  scales  compared to the otherwise random motions within the hosting system.  Such ordered fields can be  measured in  the case of planets, stars, and galaxies, or inferred indirectly by the action of their dynamical influence, such as  jets.  Whether large scale fields are amplified in situ or a remnant from previous stages of an object's history is often  debated for objects without  a definitive magnetic activity cycle.  Magnetic helicity, a measure of twist and linkage of magnetic field lines, is a unifying  tool for understanding  large scale field evolution for both mechanisms of origin. Its importance  stems from its two basic properties:  (1)  magnetic helicity is typically  better conserved  than magnetic energy; and  (2) the magnetic energy associated with a fixed amount of magnetic  helicity is minimized when the system relaxes  this helical structure to the largest scale available. Here I  discuss how magnetic helicity has come to  help us understand the saturation of and sustenance of  large scale dynamos, the need for either local or global helicity fluxes to avoid dynamo quenching, and the associated observational consequences. I  also discuss how  magnetic helicity acts as a hindrance to turbulent diffusion of large scale fields,  and thus a helper for fossil remnant large scale field origin models in some contexts.  I  briefly discuss the connection between large scale fields and accretion disk theory as well. The goal here  is to provide a conceptual primer to help the reader efficiently  penetrate the  literature. \end{abstract}


\section{Introduction}

Planets,  stars,   galaxies  are all examples of astrophysical rotators that reveal direct or indirect evidence for large scale ordered magnetic fields (Schrijver  \&  Zwaan 2000; Brandenburg \& Subramanian 2005; Shukurov 2005; Beck 2012 Roberts \& King 2013). Here ``large" scale implies a coherent flux on scales comparable to the size of the hosting rotator and most importantly, larger  than the scale of fluctuations associated with  chaotic turbulent flows.   In fact,  all of these systems show evidence for both small and large scale magnetic fields, so the fact that large scale order 
persists even amidst a high degree of smaller scale  disorder is  a  core challenge of explaining the emergence of large scale magnetic structures across such disparate classes of  rotators.  

Large scale magnetic fields are also likely fundamental to coronae and jets from accretion engines  around young and dying stars and compact objects 
(e.g. Blandford \& Payne 1982; K\"onigl 1989; Field \& Rogers 1993; Blackman et al. 2001;  Lynden-Bell 2006; Blackman \& Pessah 2009; Pudriz et al. 2012; Penna et al. 2013).
Accreting systems typically exhibit a continuum spectrum and luminosity best explained by matter accreting onto a central object
falling deeper into a potential well and thereby releasing positive kinetic energy in the form  of radiation or jet outflows. In fact, as  Fig. {\ref{bernfig1} shows,  the  classes of objects  likely harboring jets has increased  as observations have improved, and jets likely indicate the role of large scale magnetic fields.  

In young stars,  pre-planetary nebulae, micro quasars, and active galactic nuclei the jets typically  have too much collimated momenta to be driven by  mechanisms that do not involve  large scale magnetic fields (Bujarrabal et al. 2001; Pudritz et al. 2012).  From the jets of AGN, Faraday rotation from ordered helical magnetic fields is directly observed (Asada et al. 2008; Gabuzda et al. 2008; Gabuzda et al.  2012).  Because jets are anchored in the accretion engines, they play a role in extracting the angular momentum that allows remaining disc material to accrete.  The ionization fractions of accretion disks are commonly high enough, at least in regions near the very center,  to 
 be unstable to the magneto-rotational instability (MRI) (Balbus \& Hawley 1991, 1998, 2003). A plethora of numerical simulations now commonly reveal  that   the systems evolve to a nonlinear turbulent steady state whose Maxwell stress
dominates the Reynolds stress and  for which  large scale ordered magnetic  fields emerge with  cycle periods  $\sim 10$ orbit periods (e.g. Brandenburg et al. 1995; Davis et al. 2010; Simon et al. 2011; Guan \& Gammie 2011; Sorathia et al. 2012; Suzuki and Inutsuka 2013). With the caveat that  angular momentum plays a comparatively  subdominant role in structural support for stars, 
the coronae of stars and the emergence of large scale ordered fields that thread coronal holes  of the sun (e.g. Schrijver  \&  Zwaan 2000) can help shed light on  related phenomena of the rising and opening up of large scale fields that  form jets and coronae from accretion disks.
 
There are three possibilities for the  origin of large scale fields in astrophysical rotators:  The first is that the contemporary field is simply the result of advection and compression of the field that was present in the object before it was formed (e.g. Braithwaite \& Spruit 2004; Kulsrud et al. 1997;  Lovelace et a. 2009; Subramanian 2010;  Widrow et al. 2012).
  The second is that the field is in fact dynamo produced in situ, extracting free energy from shear, rotation and turbulence in such  a way as to  sustain the field against turbulent diffusion (e.g. Moffatt 1978;  Parker 1979; Ruzmaikin et al. 1988; Shukurov  2005; Charbonneau 2013).  The third possibility is  some combination of the two (e.g. Kulsrud\& Zweibel 2008).
   In systems like the Sun and Earth where cycle periods involving field reversals are observed, the need for in situ dynamo amplification is unambiguous. For galaxies or accretion disks, the  evidence for  in situ amplification of large scale fields is more indirect. 
 Regardless of  whether the fields are initially frozen in and advected, amplified in situ, or a combination of the two, the evolution of   magnetic helicity  is very helpful for  understanding the physics of magnetic field origin as we shall see.

 Much of what we can observationally infer about  magnetic fields of planets, stars, and accretion disks
  comes from information external to where the real action of magnetic field amplification and conversion of kinetic to magnetic energy occurs. Yet, most theory and simulation of astrophysical dynamos focuses on the interiors.   This  situation contrasts that of our own galaxy where we observe the field from within (e.g. van Eck et al 2011; Beck 2012), albeit on time scales too short to
  observe its dynamical evolution.   A related  point  is that  the hidden interiors of astrophysical rotators are typically flow-dominated, with the magnetic field energy density generally weaker than that of the  kinetic energy.  However in the surrounding coronae 
     of stars, accretion disks (and maybe even for galaxies) the field  dominates the kinetic energy. 
     Thus we must learn about the flow-dominated interiors from observations of the magnetically dominated exteriors and understand
     the coupling between the two.
     Laboratory plasmas of fusion devices are in fact magnetically dominated and there are many lessons learned from this contexts.   Tracking magnetic helicity evolution has proven helpful for understanding the field evolution in both magnetically dominated and flow-dominated circumstances. 

 In this paper  I discuss basic principles of magnetic helicity evolution  to  guide the physical intuition of how large scale magnetic fields arise and evolve.  This overview represents one path through the subject and is  intended as a conceptual primer to ease immersion in   
 the literature rather than a  complete detailed  review.  Many relevant  papers  will therefore regrettably go uncited---a situation that I  find increasingly difficult to avoid.
A example of an earlier  more detailed review  is Brandenburg \& Subramanian (2005).
 
In section 2,  I discuss the key physical   properties of magnetic helicity that are central to all subsequent topics discussed.
In section 3, I summarize the conceptual progress of how these principles apply to modern developments in magnetic dynamo theory. 
In section 4, I describe the simple  two-scale model  for closed systems that has come to be  useful for gaining unified insight  of large scale field evolution and dynamo saturation more quantitatively. I discuss several  applications of the two-scale theory: dynamo saturation,  the  resilience of helical fields to turbulent diffusion, and  dynamical relaxation.
In section 5, I discuss the role of helicity fluxes and their relation to the results  section 4.
In section 6, I briefly discuss the issue of gauge non-invariance of magnetic helicity.
I conclude in section 7, emphasizing the expectation that  both signs of magnetic helicity on different scales should appear
in astrophysical rotators with coronal cycle periods, and  
comment on the connections between magnetic helicity dynamics and accretion disk theory.

\section{Key Properties of  Magnetic Helicity}

Here I introduce  some basic  properties  of magnetic helicity that  underlie its role  in large scale field
generation and are essential for the sections that follow.

\subsection{Magnetic Helicity as a measure of magnetic flux linkage, twist,  or writhe}

Magnetic helicity is  defined as  volume integral of the
dot product of vector potential $\bf A$ and magnetic field ${\bf B}=\curl {\bf A}$, namely
\beq
\int \bfA\cdot \bfB  dV.
\label{1}
\eeq
To see why this is a measure of magnetic linkage (e.g. Moffatt 1978;  Berger \& Field 1984), consider two thin linked magnetic flux tubes as  shown in 
Fig. \ref{bernfig2}, with cross sectional area vectors
$d \bf S_1$ and $d\bf S_2$ respectively . Let $\Phi_1= \int \bfB_1\cdot d \bf S_1$ be the magnetic flux in tube 1, where $\bfB_1$ is the 
magnetic field.  Similarly,  for flux tube 2 we have $\Phi_2= \int \bfB_2\cdot d \bf S_2$ where $\bfB_2$ is the 
magnetic field. For both tubes,  we assume that the fields are of  constant  magnitude and parallel to $d \bf S_1$ and $d\bf S_2$ respectively. The volume integral of Eq. (\ref{1}) contributes only where there is  magnetic flux. Thus  we can split  the magnetic helicity into contributions from the two flux tube volumes to obtain
\beq
\int \bfA\cdot \bfB  dV= \int\int \bfA_1\cdot \bfB_1 dl_1 dS_1 +  \int \int\bfA_2 \cdot \bfB_2 dl_2 dS_2,
\label{link1}
\eeq
where we have factored the volume integrals into  products of line and surface integrals, with the line integrals taken along the direction parallel to $\bfB_1$ and $\bfB_2$. Since the magnitudes of $\bfB_1$ and $\bfB_2$ are constant in the tube, 
 we  can  pull $\bfB_1$ and $\bfB_2$ out of each of the two line integrals on the right of  Eq. (\ref{link1}) to write
\beq
\int \bfA\cdot \bfB  dV= \int A_1 dl_1 \int B_1dS_1 +  \int A_2  dl_2 \int B_2dS_2
= \Phi_2 \Phi_1 + \Phi_1 \Phi_2= 2\Phi_1\Phi_2,
\label{link}
\eeq
  where we have used Gauss' theorem to replace $\int A_1 dl_1=\Phi_2$, the magnetic flux of tube 2 that is linked through tube 1.
  and similarly $\int A_2 dl_1=\Phi_1$.  Note that if the tubes are not linked, then the line integral would vanish
  and there would be no magnetic helicity.
  
  Having established that the magnetic helicity measures linkage, we are poised to  understand how helicity can also  be equivalently characterized  as a measure of magnetic   twist and writhe.   Examples of  local twist and writhe are seen seen in Fig. \ref{bernfig3}, which is a roller coaster element at Cedar Point in Sandusky Ohio (USA).   The large loops each correspond to writhe, and twist is measured along the track.  (Note that ``writhe" here is what Bellan (2000) calls ``overlap" and  ``twist" here is  what  Bellan (2000) calls ``writhe" ).   
  A twisted ribbon is also shown in Fig. \ref{bernfig4},   where the amount of twist is conserved between the two panels but transferred from small to large scales.
  
 Quantitatively relating linkage to twist is nicely  accomplished  experimentally with strips of paper, a scissors and some tape (e.g. Bellan 2000).
 This is illustrated in  Fig. \ref{bernfig4.2}. The two panels show configurations with equivalent amounts of helicity as
 I now describe. Start with  a   straight strip of paper (say 20 cm long and 2 cm wide) and give the strip a full right handed twist around its long axis (by holding the bottom with your left and and twisting at the top with your right hand). Now 
 fasten the ends so that it is a twisted closed loop. This provides a  model for a  flattened, twisted closed magnetic flux tube. The result is Fig. \ref{bernfig4.2}a.  Now consider that this  tube could in fact have been composed of two adjacent flattened tubes pressed together side by side. Separating these adjacent tubes is accomplished by use of a scissors. Cut the along the center line of the strip all the way around and the result is  two linked ribbons, each of 1/2 the width of the original, and each with one right handed twist. The result is shown in Fig. \ref{bernfig4.2b}.   The use of scissors in this way has not changed the overall  helicity of the system, but has transformed it into different forms as follows:
If the original unseparated ribbon had a magnetic flux $\Phi$ then each of the 1/2 width ribbons  have  flux $\Phi/2$. From the above discussion of linkage,
the linkage of these two new ribbons gives a  contribution of helicity $2 \Phi^2/4 =   \Phi^2/2$. 
But the original uncut ribbon had a single right handed twist with total flux $\Phi$.  For the helicity of the  initial twisted ribbon to equal that of the  two linked twisted ribbons, any  twisted  ribbon must contribute a helicity equal to its flux squared.
Thus helicity is thus conserved as follows:   $\Phi^2$  is the helicity associated with the initial twisted ribbon, and this is then equal to the sum of helicity from  the linkage of the two half-thickness flux tubes $\Phi^2/2$,  plus that from the 
right handed twists in each of these two half- thickness ribbons $2 \times (\Phi/2)^2 =  \Phi^2/2$.
This conveys how both linkage and twist are  manifestations of magnetic helicity.

To see the relation between twist and writhe, consider again a straight paper ribbon. As above, give the  ribbon one
right handed twist around its long axis holding the other end fixed as in Fig. \ref{bernfig4.3}a. 
Imagine that the ends A and B are identified as the same location so it is really a closed twisted loop.
Now push the ends A and B inward toward each other and the ribbon will buckle, as seen in Fig. \ref{bernfig4.3}b.
A side view of this buckling is shown in Fig. \ref{bernfig4.3}c.  If the ends A and B are identified, then the  result is now a loop with a single unit of writhe--a large scale loop through which you can thread a rigid pole--that was derived from a single 
unit of   twist.  Thus one unit of twist helicity is equivalent to one unit of writhe helicity.

In short, one unit of twist helicity for a tube or ribbon of magnetic flux $\Phi$  is equal to one unit of writhe helicity for the same flux tube, and both are  separately equal to 1/2 of  the helicity resulting from  the  linkage of two untwisted flux tubes of flux $\Phi$.
These three different, but equivalent  ways of thinking about magnetic helicity are instructive for 
extracting the physical ideas in what follows.

\subsection{Evolution of  Magnetic Helicity}

The time evolution equation for magnetic is simply derived in magnetohydrodynamics:
The electric field is given by
\beq
\bfE=-\nabla\Phi -{1\over c}
\partial_t\bfA,
\label{1bf}
\eeq
where $\Phi$ and $\bfA$ are the scalar and vector potentials.  Using
$\bfB\cdot \partial_t \bfA= \partial_t(\bfA\cdot \bfB) +c\bfE\cdot \bfB -c\nabla \cdot (\bfA\ts \bfE)$,
where the latter two terms result  from Maxwell's equation 
\beq
\partial_t \bfB=-c\curl \bfE
\label{max}
\eeq
and the identity 
$\bfA \cdot \curl \bfE = \bfE\cdot\bfB-\nabla \cdot (\bfA \ts \bfE)$, 
we take the dot product of Eq.  (\ref{1bf}) with $\bfB$ to obtain the evolution of the  magnetic helicity density
\beq
\partial_t(\bfA\cdot\bfB)= -2c\bfE\cdot\bfB
-\div ({c\Phi}\ \bfB + c\bfE\ts \bfA).
\label{5bf}
\eeq
If we average this equation over a simply connected volume that has no boundary terms, the last terms would not contribute and we obtain
 \beq
\partial_t\lb\bfA\cdot\bfB\rb= -2c\lb\bfE\cdot\bfB\rb.
\label{6bf}
\eeq
In MHD Ohm's law is 
\beq
\bfE =-\bfv \times\bfB/c + \eta \bfJ,
\label{ohm}
\eeq
where $\bfV$ is the flow velocity and $\eta = 4\pi \nu_M /c^2$ is the resistivity for a magnetic diffusivity $\nu_M$. For ideal MHD ($\eta =0$)
 the right hand side of (\ref{6bf}) is zero, highlighting the conservation of magnetic helicity density (and thus magnetic helicity)
for a closed volume under ideal conditions and independent of the presence of velocity flows.

For comparison,  the magnetic energy density evolution, obtained by dotting Eq. (\ref{max}) with $\bfB$ and using Eq. (\ref{ohm}) is given by 
\beq
{1\over 8\pi} \partial_t\langle\bfB^2\rangle = - \eta\langle\bfJ^2\rangle
 -{1\over c }\langle {\bf v }\cdot (\bfJ\times  \bfB)\rangle,
\label{3.1.7a}
\eeq
where $\bfJ\equiv {c\over 4\pi}\curl \bfB$. 

\subsection{Magnetic Helicity is Typically Better Conserved than Magnetic Energy}

In the absence of dissipation, magnetic helicity is exactly conserved in MHD for a closed system but magnetic energy can be exchanged with a velocity field. If we  ignore the latter, when the ratio of time scale  for resistive decay of magnetic energy to  that of magnetic helicity 
is small,  magnetic helicity is more strongly conserved than magnetic energy.  To demonstrate when this  is true, we must consider that a typical astrophysical system is often turbulent, so there is not just one scale of the field but a spectrum.  The question becomes for what spectra of magnetic energy  and magnetic helicity  does the latter decay more slowly than the latter?   
Following Blackman (2004) and  working in the Coulomb gauge,  we write the (statistically or volume)  averaged magnetic energy density as
\beq
\langle\bfB^2 \rangle/8\pi=M=
\int_{k_0}^{k_{\nu_M}} M_k dk,
\label{3.1.1}
\eeq
where the brackets indicate an average,   $k_0$ and $k_{\nu_M}$ are the minimum and maximum (resistive) 
wave numbers, and  the one-dimensional magnetic energy density spectrum is given by
\beq
M_k\equiv{1\over 8\pi} \int |{\tilde {\bf B}}|^2k^2d\Omega_k  =
\int |{\tilde {\bf A}}|^2k^4d\Omega_k 
\propto k^{-q}.
\label{3.1.2}
\eeq
Here $\Omega_k$ is the solid angle in wave-number space,   $q$ is an assumed constant, 
and the tilde indicate Fourier transforms.
The magnetic helicity density
spectrum is  then
\beq
\begin{array}{r}
H_k\equiv {1\over 16\pi}\int\left[{\tilde {\bf A}}(k) {\tilde {\bf B}}^*(k) + {\tilde {\bf A}}^*(k) {\tilde {\bf B}}(k)\right]k^2d\Omega_k\\ 
= M_k f(k)/k,
\end{array}
\label{3.1.3}
\eeq
where $*$ indicates complex conjugate and  $f(k)\propto k^{-s}$ is used to define the  fraction of magnetic energy that is helical at each wave number
and $s$ is taken as a constant.
We then also have correspondingly
\beq
\langle\bfA\cdot \bfB \rangle=\int_{k_0}^{k_{\nu_M}}f(k)M_k k^{-1}dk,
\label{3.1.4}
\eeq
and the current helicity density
\beq
\langle\bfJ\cdot \bfB \rangle=\int_{k_0}^{k_{\nu_M}}f(k)k M_k dk.
\label{3.1.5}
\eeq

Now, using  Eq. (\ref{6bf}) for a closed system we have
\beq
8\pi \partial_t H=\partial_t\langle\bfA\cdot\bfB\rangle =-2{\nu_M}
 \lb \bfJ\cdot \bfB\rb,
\label{3.1.6}
\eeq
and  Eq. (\ref{3.1.7a}) gives in the absence of velocity flows,
\beq
8\pi \partial_t M=\partial_t\langle\bfB^2\rangle =-2{\nu_M}\langle(\nabla \bfB)^2\rangle.
\label{3.1.7}
\eeq
Then combining  (\ref{3.1.4}-\ref{3.1.7}),
 we then obtain
\beq
\tau_H= {-H\over \partial_t H}  =  {\int_{k_L}^{k_{\nu_M}} f(k) M_k k^{-1} dk \over 
2{\nu_M} \int_{k_L}^{k_{\nu_M}}f(k)kM_kdk}
\eeq
and
\beq
\tau_M= {-M\over (\partial_t M)_{res}} =  {\int_{k_L}^{k_{\nu_M}} M_k  dk \over 
2{\nu_M} \int_{k_L}^{k_{\nu_M}}k^2M_kdk}, 
\eeq
for the time scales of magnetic helicity and magnetic energy decay respectively.  The subscript ``{\it res}" indicates the contribution from the penultimate term in  Eq. (\ref{3.1.7})
only.  The range of  $s$ and $q$ for which 
 $R\equiv {\tau_H\over \tau_M}  > 1$ corresponds to regime in which 
 the magnetic helicity decays more slowly than the magnetic energy.
Blackman (2004) showed that $R>1$ for the combination of  $s>0$ and $3> q>0$.
and that $R<1$  for small $0 < q< 1$ and  $s <0 $. The latter range would correspond to a very unusual
circumstance in which all the magnetic helicty were piled up at small scales.  Most commonly therefore, the range for which    $R>1$   applies and magnetic helicity is typically better conserved than magnetic energy.

\subsection{Minimum Energy State of Helical Magnetic Fields}

The conclusion of the previous section---that magnetic helicity usually decays more slowly than magnetic energy-- justifies  
\textit{a posteriori} the relevance of the  question that Woltjer (1958a) considered:  If the magnetic helicity is conserved for a magnetically dominated system (ignoring velocities), what magnetic field configuration minimizes the energy?
Using a variational principle calculation, Woltjer (1958a) found that the answer is a configuration for which  $\bfJ= {c\over 4\pi}\curl \bfB =f({\bf x})\bfB$, where  $f({\bf x})$  is a scalar function that must therefore satisfy $\bfB\cdot \nabla f=0$. 
Taylor (1974,1986) considered the same question but assumed that  magnetic helicity is approximately conserved when  averaged over sufficiently large scales even in the presence of a small but finite resistive dissipation .    Essentially,  $f$ becomes a measure of the inverse gradient scale of the helical magnetic field and so minimizing the energy means decreasing $f$ as much as possible. The small amount of dissipation  
aids this relaxation via small scale reconnection as needed, such that the overall relaxed state of the field is one in which the small scale gradients smooth out to allow the gradient scales to reach the largest possible subject to the boundary conditions. This, in turn, uniquely determines  $f$ as a function of the specific boundary conditions.
(Fig. \ref{bernfig4}) shows a simple helicity conserving relaxation  process where  reconnection is not actually needed.)
The arguments of Woltjer (1958a)  and Taylor (1974)  essentially assumed that $R>1$. The previous section shows the  specific spectral conditions for this to be viable, and  solidifies the assumptions on which these results were based.

That  Woltjer (1958a) arrived at a force-free, minimum energy state under the conditions imposed-- no velocity flows, no dissipation, and a closed volume---is   evident even  without a variational calculation.
Imagine a system with initially no pressure gradients and no velocity with  a field configuration that is not force free, i.e., $\bfJ \times \bfB\ne 0$. A velocity flow will swiftly develop, violating the assumption that there is no velocity. The only way that the velocity (and thus kinetic energy) could remain zero is if the field produces no acceleration, i.e. is force free. The assumption of no velocity is therefore enough to conclude that the field must be force-free in a steady state.
Moreover, in a closed system, a force free field 
is fully helical, namely  $ \lb \bfJ\cdot \bfB\rb =-\lb \bfB\cdot \nabla^2\bfA\rb$, 
so that the 1-D Fourier spectrum of current helicity $H_c(k)$ would be   $k^2H_M(k) $ where $H_M(k)$ is the  magnetic helicity spectrum. For a fully helical system,  $kH_M(k) =M(k)= H_c(k)/k$.   Suppose the magnetic energy is predominately at  a single wavenumber $k_E$,  and we ask whether $k_E$ must increase or decrease to minimize the magnetic energy:  If  magnetic helicity is conserved then $k_EH_M(k_E)= M(k_E)$ would remain constant as $k_E$ changes. The magnetic energy  $k_EM(k_E)$ thus decreases with decreasing $k_E$. As a result, minimizing the magnetic energy for a fixed magnetic helicity would lead to as small of a $k$ as possible.  This is  the essence of the ``Taylor relaxed" state.

Note also that if we drop the assumption that  there is no velocity flow, then the momentum equation for incompressible  ($\div \bfv=0$, $\rho=constant$) flow  in the absence of microphysical dissipation is
\beq
 {\partial \bfv \over \partial t} = \rho \bfv \times (\curl \bfv) - \nabla \left(\bfv^2/2 + {P\over \rho}\right) +{\bfJ\times \bfB \over \rho}.
\eeq
Dotting with $\bfv$ and averaging over a closed volume gives 
\beq
\lb{\partial \bfv^2 / \partial t}\rb = \lb \bfv \cdot( \bfJ\times \bfB)\rb.
\label{20yes}
\eeq
In the absence of dissipation,  the right hand side of Eq. (\ref{20yes}) is  also the only contributing term to the evolution of magnetic energy in  Eq. (\ref{3.1.7a}).
Thus $\lb \bfv \cdot( \bfJ \times \bfB)\rb =0$  is the generalization  to the force free state when velocities are allowed, for it is the only way for both the magnetic and kinetic energies to remain steady.   Woltjer (1958b) extended  Woltjer (1958a) by deriving integrals of the motion for more general hydromagnetic flows and Field (1986) focused on a static extension of Woljer (1958a) to include pressure gradients.    
However, the derivation of Eq (\ref{20yes}) above  provides a simple   articulation of the generalization to the force free condition when velocity flows are allowed.


This  subsection has focused on the steady state, not the dynamical relaxation to that state.
In fact,  magnetic relaxation is a time dependent.  Even in the magnetically dominated regime of astrophysical coronae and
laboratory fusion plasmas where magnetic relaxation is considered (e.g. Bellan 2000; Ji \& Prager 2002);   this relaxation is in fact a large scale dynamo (LSD), because large scale helical  fields
grow where none were present initially, and as a consequence of helical (magnetic)  energy input on small scales. In the discussion of dynamos in the next sections, I will discuss  how 
this relaxation and the more traditional flow driven large scale field growth are different flavors within  a unified framework. Magnetic helicity evolution is fundamental to both.

\section{Helicity and  Large Scale Dynamo Saturation:  Conceptual Progress }

\subsection{Types of Dynamos and Approaches to Study Them}
Dynamos describe the growth or sustenance of magnetic fields against the otherwise competing exponential decay.
 They can be divided into two major classes:

{\bf Small Scale Dynamo (SSD)}}: This corresponds to magnetic energy amplification by turbulent velocity flows for which the dominant   magnetic energy growth occurs primarily at and below the velocity forcing scale (e.g. Kazantsev 1968; Schekochihin et al 2002,
Bhat \& Subramanian 2013; Brandenburg \& Lazarian 2013). There is a large body of work addressing the overall magnetic energy spectra of such small scale dynamo when the system is isotropically forced without kinetic helicity.  We will not focus on small scale dynamos in what follows.  

{\bf Large Scale Dynamo (LSD)}  For LSDs, the  magnetic energy grows on spatial or temporal scales larger than the dominant input scale. The input energy  can take different forms but  in all cases, the fact that the field grows on scales large
compared to those of the input energy always requires a large scale electromotive force (EMF) 
aligned with the large scale or mean magnetic  field, such that $(\overline{ \bfv\times \bfb})\cdot\bbB\ne 0$,
where the overbars indicate a  spatial, temporal, or ensemble average and  $\bfv$ and $\bfb$ are the velocity magnetic fields associated with turbulent  fluctuations when the  total velocity and magnetic field are written as sums of fluctuations plus mean values as  
 $\bf V= \bfv +\bbB$ and  $\bf  B=\bfb+\bbB$ respectively.
This implies a  source of large scale magnetic helicity is involved in field amplification as we shall see more explicitly later.

Two sub-classes of LSDs can be distinguished, based on the nature of the dominant energy input:

{\it 1. Flow Dominated}:   For this subclass,  the input energy is  kinetic energy dominated and the EMF can be sustained, for example, by kinetic helicity, as  in the classic Parker-type solar dynamo and its extensions  (e.g. Moffatt 1978; Parker 1979; Krause \& R\"adler 1980; Pouquet et al. 1976; Blackman \& Field 2002; Blackman \& Brandenburg 2002). Alternatively, there is emerging agreement that  a  combination of shear and fluctuating kinetic helicity can conspire to produce an LSD  (Vishniac \& Brandenburg 1997; Brandenburg 1995;  Brandenburg 2005; Yousef et al. 2008; Heinemann et al. 2011; Mitra \& Brandenburg 2012; Sridhar \& Singh 2013). The EMF can  also be sustained by  magnetic  helicity fluxes, either local (within sectors of a closed volume) or global (fluxes thought the system boundary) as we will later discuss.

{\it 2. Magnetic Relaxation LSD}: This subclass of LSDs occurs when a system is initially magnetic energy dominated and the EMF is sustained by magnetically dominated quantities.  Typically,  injection of small scale magnetic helicity  drives instabilities that facilitate the relaxation of the system and transfer of the magnetic helicity to  large scales. It is a dynamo because field grows on large scales where there was little  initially, and the field on large scales is sustained against decay.    In plasma fusion devices where such MRD occur, helicity fluxes  are the key sustainer of the EMF (e.g. Strauss 1985, 1986; Bhattacharjee \& Hameri 1986;  Ortolani \& Schnack 1993; Ji et al. 1995;  Bellan 2000; Ji \&  Prager 2002)
In astrophysics such MR-LSDs likely occur in coronae, where the field is injected from below on small scales relative to the corona and further relaxation occurs. Astrophysical rotators with coronae likely have flow driven dynamos in the interior,
  coupled to MRDs in coronae. Ironically, for the sun and stars, the observed field measured is from the base of the corona outward. Thus  we can directly observe the MR-LSD processes better than the interior  flow driven LSD, although the two are dynamically coupled.

In studying  dynamos there are different  approaches used depending on the goal.
One approach is the ``kitchen sink" approach, that is, try to perform quasi-realistic numerical experiments to match observational features
with as realistic of  conditions  as possible given numerical limitations (e.g. Glatzmaier 2002).  
A second approach is to carry out semi-empirical model calculations based on linear theories but with dynamo transport coefficients
empirically tuned to match general observational features and cycle periods (e,g, Dikpati \& Gilman 2009; Charbonneau 2013), and without going after the physics of nonlinear quenching. A third approach, and the one most reviewed in  this primer,  is identifying basic principles of 
 EMF sourcing and quenching.   This  involves pursuit of a ``first principles"  mean field theory and  comparison to minimalist simulations to test the predictions of the theory. 
The  goal is to  develop a mean field  theory that isolates the key physics from the nonlinear mess and eventually use the insights gained to  inform more detailed models.

\subsection{20th Century vs. 21st Century Dynamos: Physical Role of Magnetic Helicity Conservation}

The 20th century textbook mean field dynamo theory (MFDT) of standard textbooks \cite{moffatt,parker,kr80}
 is a  practical approach to modeling LSDs in turbulent rotators. 
 These approaches  focused  mostly on initially globally reflection asymmetric rotators where the 
the EMF  is sustained  by   kinetic helicity.
But for $\sim 50$ years, this    theory
 lacked a  saturation theory to predict how strong the large scale fields get
before  quenching  via the back-reaction of the field on the driving flow.

In fact, the inability of the 20th century textbook theories to predict mean field dynamo saturation arises  because 
they  fail to conserve magnetic helicity.  
This is illustrated in a most minimalist way in the  diagrammatic representations of  Fig \ref{bernfig5}  for the so called $\alpha^2$ dynamo, a dynamo that just depends on small
scale helical velocity motions (discussed more quantitatively later).  The figure shows that from an initial toroidal  loop, 
 the four small scale helical eddies, each with the same sign of  kinetic helicity $\bfv\cdot\curl \bfv$
make four small poloidal loops as shown.  The result is a net toroidal EMF  $\emf_\phi =\lb v_z b_r \rb $ which has the same sign inside the loops of all the eddies whether they moved up or down.
This in turn leads to two net large scale poloidal loops (shown in blue).  
But recall that magnetic helicity is   a measure of linkage. The first panel has only   a single red loop, but seemingly evolves to a configuration with the red loop linked to the
two blue loops.  That is, the mean field has gone from no helicity, to two units of linkage helicity. This  lack of helicity conservation highlights  an unphysical feature of 20th century LSD theory.

The commonly used schematics and diagrams of more finely tuned and observationally relevant solar dynamo models, such as flux-transport models \cite{dikpati09}, also do not conserve magnetic helicity for essentially the  same reason as in Fig. \ref{bernfig5}.
There is typically  a step where the large scale field gains helicity when a  poloidal loop emerges from  an initially toroidal field via
the action of the Coriolis force.  Flux transport dynamos are otherwise impressive in that they can be tuned to 
agree with large scale  observations of  the solar magnetic field  (e.g. Wang \& Sheeley 2003), but the fact that  certain parameters must be tuned by hand (rather than derived from first principles) highlights  that magnetic helicity dynamics have not yet been incorporated.

How do we  reconcile LSDs  with the conservation of magnetic  helicity?
The  answer is shown in Fig. \ref{bernfig7} (figure from Blackman \& Hubbard  2014).  There the same set of diagrams as in Fig. \ref{bernfig5} are shown with the field lines
represented by ribbons or flux bundles instead of lines.
Now conservation of helicity is maintained: the writhe of the loops in the second panel is compensated for by the opposite sign of twist helicity along the loops. The linkage that results in the third panel for the large scale field is compensated for by
the exact opposite amount of small scale twist helicity along the loops.  
This conservation of magnetic helicity  of an untwisted loop subjected to writhe can be demonstrated by using an ordinary belt.   

The diagrammatic solution to the missing helicity problem  is also the key to understanding  nonlinear quenching. In a system driven with kinetic helicity, the back reaction on the driving flow comes primarily from the buildup of the small scale twist. Field lines with ever increasing small scale twists become harder to bend.    It is  this small scale twist  that produces a small scale current $\bfj\equiv \bfJ-\bbJ$ (where $\bbJ=\curl \bbB$) and a Lorentz force when coupled to the mean field,   $\bfj\times \bbB$, that produces the back reaction against the small scale flow.
The driving kinetic helicity acts as a pump,  segregating the signs of the magnetic helicity between scales.
but the  presence of small scale twist eventually quenches the dynamo.

 How catastrophic this quenching is, depends  on the specific
type of dynamo: For the $\alpha^2$ dynamo (a dynamo without differential rotation), the quenching slows the growth to resistively limited rates after an initial fast growth phase, but the field does not decay (Field \& Blackman 2002; Blackman \& Field 2002).
In contrast, for the $\alpha-\Omega$ dynamo (i.e. a dynamo for which the  toroidal field growth is dominated by shear from differential rotation), the quenching is such that the field starts to grow and then decays rapidly (e.g. Shukurov et al. 2006, Sur et al. 2007).  
That helicity fluxes might alleviate quenching   and   sustain  the EMF 
(e.g. Blackman \& Field 2000a Vishniac \& Cho 2001; Brandenburg \& Subramanian 2005; Shukurov et al. 2006; Ebhami \& Bhattacharjee 2014) has emerged as the most plausible way around this decay.
Because of the way our understanding of the quenching has developed, 
the  role  of magnetic helicity flux is often expressed as  a solution to the quenching problem, however the helicity flux could have been the driver of the growth from the start (as in laboratory plasma dynamos, e.g. Strauss 1985; 1986; Bhattacharhee \& Hameri 1986) and thus the issue of quenching would not  arise in the discussion in the first place.  This is something to keep in mind in reading the  literature
across subfields. 


The physical description described in the previous subsection  is supported by detailed calculations.
Mathematically coupling the dynamical evolution of magnetic helicity into the dynamo equations can largely explain the saturation seen in simulations. The  connection between magnetic helicity and large scale dynamos was first evident in the   spectral model of helical MHD turbulence of Pouquet et al. (1976).  They demonstrated  an inverse transfer growth of large scale magnetic helicity  for which the   driver is the difference between kinetic and current helicites.
In Kleeorin \& Ruzmaikin (1982)  an equation that couples the small scale magnetic helicity to the mean electromotive force is present, but the time evolution was not studied.  The spectral work of Pouquet et al. (1976) was  re-interpreted (Field \& Blackman 2002) and re-derived (Blackman \& Field 202) as a  Ôuser-friendlyÕ time-dependent mean field theory
exemplified for the $\alpha^2$ dynamo case. Indeed it was found that 
 $ 
  \lb\bfv\cdot \curl \bfv\rb$ grows large scale mag. helicity of one sign and small scale mag. helicity of opposite sign.   The latter quenches LSD growth and  matches simulations of Brandenburg (2001) and  subsequent papers.
A similar framework shows that large scale helical fields are resilient to turbulent diffusion (Blackman \& Subramanian 2013; Bhat et al. 2013) 
 These developments  will be discussed later in section 4.

The study of magnetic helicity dynamics in systems unstable to the  magneto rotational   instability (MRI)   is also emerging (e.g. Vishniac 2009; Gressel 2010, K\"apyl\"a \& Korpi 2011; Ebrahimi \& Bhattacharjee 2014): MRI simulations exhibit LSDs driven by an EMF sustained by something other than kinetic helicity. e.g. current helicity or helicity fluxes.   Helicity fluxes may sustain EMF and alleviate premature quenching in realistic systems with global boundaries OR in sub-volume local sectors even if the system is closed (e.g. vertically periodic shearing box)
This will be addressed in Section 5.

\section{Two-Scale Approach to Helicity Dynamics Provides Physical Insight}

We now summarize the derivation of the  two-scale equations for  magnetic helicity evolution from which the above
physical discussion arises. We will then discuss two applications of these equations, one for the $\alpha^2$ dynamo and the second for the
resilience of helical fields to decay. A third application to dynamical relaxation is briefly mentioned as well.

\subsection{Basic Equations}

We follow standard procedures   \citep{2002PhRvL..89z5007B,2005PhR...417....1B,BS13} and 
break each variable into large scale quantities (indicated by an overbar) and fluctuating quantities (indicated by lower case).  
  We indicate global averages by brackets.  The overbar indicates a more local average than  brackets,  (and can also be  an average over reduced dimensions)  but still over a large enough scale such that  both local and global averages of fluctuating quantities vanish.
 
 The analogous procedure that led to Eq. (\ref{5bf}) for the evolution of the total magnetic helicity, leads to separate expressions for the contributions to magnetic helicity from the large and small scale given by 
\beq
{1\over 2}\partial_t\lb\overline{\bfa\cdot\bfb}\rb=
-\lb\emfb\cdot\bbB\rb
-\nu_M \lb\overline{ \bfb\cdot\curl\bfb}\rb-{1\over 2}\div(c\overline{{\phi} \bfb} + c\overline{\bfe\ts \bfa}),
\label{5aa}
\ee
and
\beq
{1\over 2}\partial_t\lb\bbA\cdot\bbB\rb=\lb\emfb\cdot\bbB\rb-\nu_M\lb\bbB\cdot\curl\bbB\rb-{1\over 2}\div ({c\overline\Phi}\ \bbB + c\bbE\ts \bbA).
\label{6aa}
\eeq
where $\emfb\equiv \overline{\bfv\ts\bfb}=-\bbE +\eta \bbJ= -{\bf E}+ \bfe + \eta({\bf J}  -\bfj)  $ is the turbulent electromotive
force.
The simplest expression for $\emfb$ that connects 20th century dynamo theory to 21st century 
 makes use the 'tau'   or   `minimal tau'  closure approach for incompressible MHD \citet{2002PhRvL..89z5007B,2005PhR...417....1B}. This  means replacing  triple correlations by a damping term on the grounds that   the EMF $\emfb$ should decay in the absence of $\bB$.  The result is  
\beq
\partial_t \emfb = \overline{ \partial_t \bfv  \times \bfb} + \overline{ \bfv \times \partial_t \bfb }
={\alpha \over {\tau}}\bbB-{\beta\over {\tau}}\curl\bbB-\emfb/{\tau},
\label{7aa}
\eeq
where ${\tau}$ is a damping time 
and 
\beq
\alpha \equiv {{\tau}\over 3} \left({\lb \bfb\cdot\curl\bfb \rb\over 4\pi\rho}- 
\lb \bfv\cdot\curl\bfv \rb\right) \quad  {\rm and} \quad 
\beta\equiv   {{\tau} \over 3}\lb v^2\rb,
\label{7aab}
\eeq
and we assume $\lb\bfv\cdot\curl\bfv\rb \simeq\overline{\bfv\cdot\curl\bfv}$ and $\lb\bfv^2\rb \simeq\overline{\bfv^2}$.

Keeping the time evolution of  $\emfb$  as a separate equation to couple
into the theory and solve allows for oscillations and phase delays between extrema of field strength and extrema of
$\emfb$ (Blackman \& Field 2002) and 
 are of observational relevance (as they have  been used to explain the phase shift between
spiral arms and dominant large scale mean field magnetic polarization in the galaxy (Chamandy et al. 2013).
But  simulations of magnetic field evolution in the simplest forced  isotropic helical turbulence 
 reveal that a good match to the large scale magnetic field evolution
can be achieved even when the left side of Eq. (\ref{7aa}) is ignored  and 
 $\tau$
 is taken as the eddy turnover time associated with the forcing scale. 
We adopt that  approximation here and so Eq. (\ref{7aa}) then gives
\beq
\emfb =\alpha \bbB-\beta \curl\bbB.
\label{emfb}
\eeq
 Eqs. (\ref{5aa}) and (\ref{6aa}) then become
\beq
{1\over 2}\partial_t\lb\overline{\bfa\cdot\bfb}\rb=
-\alpha\lb\bB^2\rb +\beta\lb\bbB\cdot\curl\bbB\rb
-\nu_M \lb\overline{ \bfb\cdot\curl\bfb}\rb
\label{5ab}
\eeq
and
\beq
{1\over 2}\partial_t\lb\bbA\cdot\bbB\rb=
\alpha\lb\bB^2\rb -\beta\lb\bbB\cdot\curl\bbB\rb
- \nu_M\lb\bbB\cdot\curl\bbB\rb.
\label{6ab}
\eeq
The  energy associated with the small scale magnetic field does not enter  (Gruzinov \& Diamond 1996) $\emfb$  so 
it does not  enter Eqs. (\ref{5ab}) and (\ref{6ab}). 
 It arises as a higher order hyper diffusion correction (Subramanian 2003)  which  we  ignore.
However,   the energy density  in the large scale field  $\propto \bB^2$ {\it does} enter Eqs.  (\ref{5ab}) and (\ref{6ab}). In general we need  a separate 
equation for the energy associated with the energy of the mean field.  Fortunately, for the simplest $\alpha^2$ dynamo  discussed in Sec. 4.2 below, the non-helical large scale field does not grow even when the additional equation is added.
And for the  decay problem of section 4.3, the non helical part of the magnetic energy decays very rapidly. As such, it is acceptable  to assume that the large scale field is fully helical for present purposes.

The essential implications of the coupled  Eqs. (\ref{5ab}) and (\ref{6ab}) for a closed or periodic system 
are  revealed in standard approaches 
where the  large scale overbarred  mean magnetic  quantities are now indicated with  subscript ``1", small scale  quantities by subscript ``2". The kinetic  forcing scale is indicated by  subscript $f$.
In the usual two-scale model for the $\alpha^2$ dynamo,  the kinetic forcing wavenumber $k_f$ is assumed to be  same as that for the small scale magnetic  fluctuations $k_2$. Relaxing this provides some versatility, but we  take $k_f=k_2$  here for simplicity.
 We assume that  the  wave number  $k_1$    associated with the spatial variation scale of large scale quantities  satisfies  $k_1\ll k_2$, where $k_2$ is the wave number  associated with  small  scale  quantities.     
Applying these   approximations  to a closed or periodic system,  we   then  use
 $\lb \bbB \cdot \curl \bbB \rb = k_1^2 \lb \bbA \cdot \bbB \rb\equiv k_1^2 H_1$,  $\lb \bbB^2 \rb = k_1 |H_1|$,
         along with   
   $\lb \bfb\cdot \curl \bfb \rb = k_2^2 \lb \bfa \cdot \bfb \rb=k_2^2 H_2$. We will assume that $H_1\ge 0$ for the example cases studied below.
 
Non-dimensionalising by  scaling lengths in units of $k_2^{-1}$,
and time in units of $\tau$, where we assume $\tau=(k_2 v_2)^{-1}$,
we have 
\[h_1\equiv {k_2 H_1\over 4\pi \rho v_22}, \
h_2\equiv {k_2 H_2\over 4\pi \rho v_2^2}, \
h_v\equiv {H_V\over  k_2v_2^2}, \
R_M\equiv {v_2\over \nu_M k_2}.
\]
With these non-dimensional quantities, Eqs. (\ref{5ab}) and (\ref{6ab}) become
\beq
\partial_\tau h_1 = {2\over 3}\left(
h_2-h_v\right) {k_1\over k_2}h_1-{2\over 3}\left({k_1\over k_2}\right)^2 h_1-{2\over R_M}\left({k_1\over k_2}\right)^2h_1,
\label{9}
\eeq
\beq
\partial_\tau h_2 = -{2\over 3}\left(
h_2-h_v\right){k_1\over k_2}h_1+{2\over 3}\left({k_1\over k_2}\right)^2 h_1-{2\over R_M}
h_2,
\label{10}
\eeq
Eqs.  (\ref{9}) and (\ref{10})  comprise a powerful set  for capturing basic helicity dynamics for closed or periodic
volume low lowest order in turbulent anisotropy. 
They help to conceptually unify  a range of physical process depending on the initial conditions. Examples are described in the next subsections.

\subsection{Large Scale Field Growth: the $\alpha^2$ Dynamo Example}

Consider a closed  or periodic system with an initially weak seed  large scale magnetic helicity at wave number  $k_1$, with initially $0<h_1(\tau=0) \ll 1$, and with $h_2(\tau =0)=0$.    Suppose the system is steadily forced isotropically with kinetic  helicity at wave number $k_2\gg k_1$ such that $h_v=-1$ in
Eqs.  (\ref{9}) and (\ref{10}).     Solutions to this problem are shown for different time ranges as the solid lines  in Fig. {\ref{bernfig8}a and  Fig. {\ref{bernfig8}b  for $k_2/k_1=5$ and three different magnetic Reynolds numbers (as  in Field \& Blackman 2002 rather than 
Blackman \& Field 2002 since $\partial_t\emfb$ is ignored.)

The  basic interpretation of the curves is this:  At early times, the $R_M$ dependent terms in Eqs.  (\ref{9}) and (\ref{10}) are small and $h_1$ grows exponentially with a growth rate $\gamma= {2k_1\over 3k_2}\left(
h_v-{k_1\over k_2}\right)$ from Eq. (\ref{9}).
The first three terms on the right of Eq. (\ref{10}) have the same magnitude but opposite sign as those on the right of Eq.
(\ref{9}), so $h_2$ grows with opposite sign as $h_1$ with the same growth rate during this kinematic, $R_M$ independent growth phase.  
This phase lasts  until the compensating growth of $h_2$ 
 becomes large enough   to significantly offset the driving from $h_v$ and reduce $\gamma$ to a level for which the $R_M$
 terms become influential. The $R_M$ term of Eq. (\ref{9}) is $k_2^2/k_1^2$ times  that Eq. (\ref{10}), so  $h_2$ is 
 quenched by its $R_M$ term earlier,  allowing $h_1$ to
 continue growing, albeit now at an  $R_M$ dependent rate. In the $R_M\gg 1$ limit, the growth rate past the initial $R_M$ independent regime is extremely small and generally not astrophysically relevant.
 
At the end of the kinematic phase,  it can be shown analytically  that  the energy in the helical field  
grows to $B_1^2\sim k_1H_1 = {k_1\over k_2}\left(1-{k_1\over k_2}\right) v_2^2$. At this point the three different $R_M$ curves in Fig. \ref{bernfig8}b
diverge with the lowest $R_M$ curve being the faster to reach the asymptotic steady state.
By setting the left sides of  Eqs. (\ref{9}) and (\ref{10}) to zero, the 
the asymptotic saturation value can be shown to be 
$B_1^2=k_1H_1= {k_2\over k_1}\left(1-{k_1\over k_2}\right) v_2^2$, or $h_1=20$ for $k_2=5k_1$ as seen in Fig. \ref{bernfig8}a.
The final value of $h_1$ is independent of $R_M$ even though the time to get there is longer for larger  $R_M$.

In  Fig. \ref{bernfig8}a, the dotted  lines represent 
an $R_M$ dependent empirical fit formula  to the late time data in numerical simulations of Brandenburg (2001) of the $\alpha^2$ dynamo.
The agreement between the  solutions to equations (\ref{9}) and (\ref{10})  and  the empirical fit formula to the simulations is quite good at late times. The dotted lines Fig. \ref{bernfig8}b represent the {\it  artificial extension}  of this empirical fit formula
 to early times where it  does not apply.
The  asymptotic regime has an $R_M$ dependence
 whereas the early time regime does not.  More recent simulations   (Brandenburg 2009) have confirmed that the  large scale growth rate varies by only 16\% in the kinematic regime when $R_M$ varies by a factor of 100.
 The success of these two simple equations in capturing saturation features of dynamo simulations highlights the importance
 of coupling magnetic helicity into the dynamics of field growth.
 
   Examples of the large scale  magnetic and kinetic energy spectra for the $\alpha^2$ dynamo
 from direct numerical simulations in a periodic box starting with an initial weak seed field and forced with helical forcing at $k=5$ 
 are shown in Fig. 9.  The left panel shows a $64^3$ simulation and the saturated end state of magnetic and kinetic energies 
 for different forcing fractions of kinetic helicity 
 $f_h= |\lb\bfv\cdot \curl \bfv\rb|/ k_f \lb \bfv^2\rb$ and magnetic Prandtl number =3 from Maron \& Blackman (2002) .
 In the left panel, the thick red and blue lines are the magnetic and kinetic energy spectra  for the case of $f_h=1$. The thin
 red and blue lines are the magnetic and kinetic energy spectra  for the case of $f_h=0$.
  The right panel is a $512^3$ simulation from Brandenburg et al. (2012) for $f_h=1$ for unit magnetic Prandtl number.  
 In the right panel, blue indicates kinetic energy spectrum and red indicates magnetic energy spectrum.
The thick and thin red lines of the left panel thus correspond respectively to the red and blue lines of the right panel in that 
these are all for the case of $f_h=1$.    Both panels  show the dramatic emergence of the large scale $k=1$ field for helical forcing.
The thick blue curve in the left panel shows the absence of the large scale $k=1$  magnetic field for $f_h=0$



\subsection{Helical Field Decay}


A second use of  Eqs. (\ref{9}) and (\ref{10}) is to study how helical fields decay.
As alluded to earlier, when a magnetic activity cycle period with field reversals is evident, the need for an in situ LSD is unambiguous.
But when no cycle period is detected, the question of whether otherwise inferred large scale fields 
are LSD produced or merely a fossil field often arises.   The question of how efficiently magnetic fields decay
in the presence of turbulence is important  because if a  fossil field would have to survive this diffusion to avoid the need for an in situ dynamo. 

Most work on the diffusion of large scale fields has not distinguished between the diffusion of helical vs. non-helical  large scale fields.
  Yousef et al. (2003) and  Kemel et al. (2011)  found  that that fully helical large scale fields 
 decay more slowly than  non-helical large scale fields in numerical  simulations. 
 Blackman \& Submramanian (2013, hereafter BS13) analyzed  (\ref{9}) and (\ref{10}) for the field decay problem
 and identified a critical  helical large scale magnetic energy above which decay is slow
 and below which decay is fast when the small scale helicity $h_2$  is initially zero.
  Bhat et al (2014) further developed this theory and tested the results with numerical simulations   and emphasized a distinction between two problems: the case studied by BS13  and the case in which the initial field decays slowly and then it transitions to fast decay .
 
The basic result in BS13 is captured by the solution  to  Eqs. (\ref{9}) and (\ref{10}) for  initial conditions for which $h_1>0$, $h_2=0$, $h_v=0$. This corresponds to a case in which  there is no kinetic helicity, just a driving turbulent kinetic energy that causes turbulent diffusion of $h_1$  through the penultimate term on the right of Eq. (\ref{9}).  The question studied is how does $h_1$ evolve for these circumstances as a function of its initial value? The solutions  are shown in  Fig. {\ref{bernfig10})  adapted form  Bhat et al. (2014).
 From top to  bottom the curves  correspond to  the initial values  of large scale helical field energy $k_1h_{1,0}/k_2$ in units of equipartition with the turbulent kinetic energy.
For the chosen ratio $k_2/k_1=5$, the value $k_1h_{1,0}/k_2=0.04=(k_1/k_2)^2$ and  marks a threshold initial value below which $k_1h_1/k_2$ decays  rapidly via turbulent diffusion and above which it  decays restively slowly (actually, at  twice the resistive diffusion rate).  Since   $k_1h_1/k_2$ is   the dimensionless magnetic energy associated with $h_1$, 
 the energy in the helical field need only be  at least $k_1^2/k_2^2$ times that of the turbulent kinetic energy to avoid fast decay. 
The   implication is that helical field energy above
a modest value decays resistively slowly even in the presence of turbulent diffusion. This contrasts the behavior of the non-helical
part of the large scale magnetic, which always decays at the turbulent diffusion rate independent of the presence or absence
of a helical component (BS13).

Why should the helical field resist turbulent diffusion and what determines the critical value?
The answer is as follows (BS13; Bhat et al. 2014):
Slow decay of $h_1$ occurs when  the last term on the right of (\ref{9}) is  
no  smaller than  the sum of the  first two terms on the right ($h_v=0$ for the present case). 
For large $R_M$, each of those first  two terms is separately much larger than the last term
so their combination would have to nearly cancel to meet the aforementioned condition. 
These same terms also appear in the equation for $h_2$ with opposite sign. 
Initially,  $h_1>0$ and $h_2=0$ and if we seek the initial value of $h_1$ for slow decay, 
we note that  slow decay can only occur after a very rapid 
evolutionary phase (with negligible dependence on  $R_M$) where a swift buildup of $h_2$
leads to an approximate balance between these aforementioned two terms. 
The needed  amount of $h_2$ to abate decay of $h_1$ can be estimated by  balancing first two terms on the right of  either   (\ref{9}) or (\ref{10}) for our case of $h_v=0$. This  gives 
$ h_2  \sim (k_1/k_2)$.
But since the only  source of $h_2$  is $h_1$, this  value of $h_2$ is also the 
minimum value of the initial  $h_{1,0}$  needed  for  slow  decay of $h_1$.  That is, $h_{1,0}>h_c\equiv  k_1/k_2$ for  slow decay.

  A qualitative comparison between
the numerical simulations and the analytic solutions for the case just described is shown in Fig 9b. 
The behavior in the simulations looks similar to that of the analytic model. The one caveat is that the simulations do not achieve the resolution needed to ensure that the last term of  Eq. (\ref{6ab}) is smaller than than the penultimate term for $h_1$ near the small critical value of $k_1/k_2$.    However, as mentioned above,  in addition to the case 
just described where the threshold initial value of $h_1$ for slow decay is sought, Bhat et al. (2014) also studied the 
the case for which the field is initially above the threshold for slow decay, and later transitions to fast decay.  In that case they show that  the value of $h_1$ at the transition is independent of  $k_1/k_2$, unlike the case discussed above.
In this second case, $h_2$ has already saturated by the time $h_1$ makes the transition to fast decay and so the estimate of the  threshold of $h_2$ (and thus $h_1$) for the previous case does not apply.  
 This result and the distinction between the two cases are both contained within the analytic framework of Eqs. (\ref{9}) and (\ref{10}) as discussed in Bhat et et al. (2014) where  theory and simulation are shown to agree.   Confidence in the overall theory and physical interpretation of both cases is  bolstered by this correspondence.

Taken at face value, the survival of   helical fields to turbulent diffusion
 may provide  rejuvenated credence to  pre-galactic mechanisms of  large scale field production
  that produce  sufficiently strong  helical fields (Field \& Carroll 2000; Copi et al. 2008; D\'iaz-Gil et al 2008; Semikoz et al. 2012, Tevzadze et al. 2012, Kahniashvili et al. 2013)
because such helical fields could then avoid decay by supernova driven turbulent diffusion over a galactic lifetime in  the absence of boundary terms.  Although most energy in large scale galactic magnetic fields resides in non-helical toroidal fields,  as long as the turbulent decay time for the non-helical field exceeds the linear shear time, we can expect a predominance of non-helical field in a steady state, even without an in situ dynamo to regenerate the poloidal fields: The helical field provides a minimum value below which the toroidal field cannot drop. The toroidal field  enhancement over the poloidal field would be  that which can be linearly shear amplified in a non-helical field diffusion time.     Similar considerations regarding the survival of helical fields  could apply  for the large scale fields of stars and accretion disks. 

Another implication of slow diffusion of helical fields is that 
the observation of a helical large scale field in jets from Faraday rotation  (Asada et al. 2008; Gabuzda et al. 2008; Gabuzda et a. 2012) 
 does not guarantee magnetic energy domination (Lyutikov et al. 2005)  on the observed scales (BS13). 


The calculations  just discussed  do not include  buoyancy or other boundary loss terms that could extract  large scale helicity at a rate that may still need to be re-supplied  from within  the rotator.  If such terms are important, then both helical and non-helical large scale fields would deplete, and an  in situ dynamo would be needed for replenishment.  But this shifts the focus from turbulent diffusion  to that of boundary loss terms in assessing the   necessity of in situ dynamos. More on flux and boundary terms will be discussed in the next section.

\subsection{Dynamical Magnetic Relaxation}

The resilience of helical fields to  turbulent diffusion of the previous section 
is actually the result of  the current helicity part of the $\alpha$ effect in the language of dynamo
theory rather than  an intrinsic change of  the diffusion coefficient $\beta$.  
In this respect, the resilience of large scale helical field to decay is very rooted in the 
basic  principles of Sec 2.   Namely,  magnetic helicity has the lowest energy
when on the largest scale.  Diffusing it to small scales while conserving magnetic helicity  is  fighting against this  relaxation. 

In fact we can also use Eq. (\ref{9}) and Eq. (\ref{10}) to study yet a different problem. Starting with $h_1\ll1$ and $h_2 \gg h_1$,
and $h_v=0$ we can solve for the evolution of $h_1$. Indeed, Blackman \& Field (2004). Kemel et al. (2011), and  Park \& Blackman (2012)  carried out such calculations using versions of Eqs. (\ref{9}) and (\ref{10})
where the system is initially driven with $h_2$  and found that it does indeed capture the dynamical relaxation of magnetic helicity
 to large scales: $h_1$ grows exponentially as the helicity is transferred from $h_2$. In this case, the large scale field grows with the same
 sign as $h_2$, and the combination of the turbulent diffusion and a modestly growing $h_v$ emerge as the back reactors, contrasting the $\alpha^2$ dynamo case of Sec. 4.2 above.

Although the dynamics of magnetic helicity and dynamical relaxation have often not been explicitly discussed in the context of fossil field origin models of stars   \cite{2004Natur.431..819B}, the same basic principles are relevant.

\section{Helicity Fluxes}

The previous section did not include the role of flux  or boundary terms in the evolution equations.
While the role of helicity fluxes in sustaining magnetic relaxation dynamos in laboratory fusion plasmas has  been long-studied as essential (Strauss 1985; 1986; Bhattcharjee \& Hameri 1986; Bellan 2000)  the awareness of its importance for  astrophysical contexts has  emerged more recently
(Blackman \& Field 2000ab; Vishniac \& Cho 2001; Blackman 2003;  Shukurov et al. 2006; K\"apyl\"a et al. 2008; Ebramhimi \& Bhattacharjee 2014)

Laboratory plasma helical dynamos in a reversed field pinch (RFP, Ji \& Prager 2002)  for example, 
 typically involve a magnetically dominated initial state with a dominant mean magnetic toroidal 
magnetic field. When an external toroidal electric field is applied along this torioidal field, 
a current is driven along the magnetic field which injects magnetic helicity
of one sign on small scales. This generates a poloidal  field.
For sufficiently strong applied electric fields, the system is driven
 far enough from its relaxed state that 
helical tearing or kink mode instabilities occur.  The consequent 
fluctuations produce a turbulent EMF $\emfb$ that drives the system back toward the relaxed state.  
As  discussed in section 2,  the relaxed state is the state in which the magnetic helicity
is at the largest scale possible, subject to boundary conditions.
When the  helicity injection is externally sustained, a dynamical equilibrium with oscillations can incur as  the system evolves toward and away from the relaxed state. The time-averaged  
$\emfb$ is maintained by a spatial (radial) flux of small scale 
magnetic helicity within the plasma.  The injection of helicity is balanced by the dynamo relaxation,
so the  dynamo sustains the large scale field configuration  against decay.

The simplest circumstance revealing the importance of flux terms is
  a steady state for which  the left hand side of Eqs. (\ref{5aa}) and (\ref{6aa}) vanish. 
  Then, if divergences do not vanish,
the magnetic field aligned EMF would be sustained by helicity fluxes, whose divergences are equal and opposite for the large and small scale contributions,
that is 
\beq
0=2\lb\emfb\cdot\bbB\rb 
-c\div\lb{\overline\Phi} \bbB + \bbE\ts \bbA\rb
\label{26}
\ee
and 
\beq
0=-2\lb\emfb\cdot\bbB\rb-c\div\lb{\phi} \bfb + \bfe\ts \bfa\rb.
\label{27}
\ee
Combining these two  equations reveals that  the divergences 
of large and small scale helicity through the system
are equal in magnitude but oppositely signed.

The specific observational interpretation of  flux divergence terms depends on   the averaging procedure.
If the averaging is taken over the entire interior of a rotator such as a star or disk, then such non-zero fluxes in a steady-state would imply  equal  magnitude but oppositely  signed   rates of large and small scale magnetic helicity flow  through the boundary  into the corona (Blackman \& Field 2000b).  In a steady state,  each hemisphere would receive both signs of magnetic helicity
but the respective signs on large and small scales would be reversed in the two hemispheres.   

Complementarily,  Blackman (2003, Figure 9 therein)  showed  how  an imposed  preferential small scale helicity flux  could reduce quenching in the $\alpha^2$ dynamo and the result is shown  in Fig. \ref{bernfig11a} to a larger value of $R_M$.  A simple term of the form $-\lambda h_2$ was added to Eqn (\ref{10}) to make these plots and solutions for different values of $\lambda$ are shown. The left panel shows the late time saturation value is increased with increasing $\lambda$ and the right panel shows the tendency that for large enough $\lambda$ the kinematic regime (the regime independent of $R_M$) can be extended.  Del Sordo et al. (2013) have studied  numerically  the relative role of advective  and diffusive fluxes.  Their generalized $\alpha^2$ dynamos can become oscillatory with even a weak advective wind,  due to the spatial dependence of the imposed kinetic helicity.  They do indeed find that for $R_M>1000$ the advective flux dominates the diffusive flux and helps alleviate the resistive quenching. However predicting analytically the exact value of the critical $R_M$ where this occurs requires further work.

\subsection{Helicity Fluxes in  Galactic and Stellar Contexts}

The $\alpha^2$ dynamo has no shear but helicity flux in the presence of  shear is particularly important   because generalizations of Eqs.  (\ref{9}) and (\ref{9}) to the  $\alpha-\Omega$ LSD otherwise lead to large scale field  decay. Fig. \ref{bernfig11a} can be contrasted    in this regard.
 Fig, \ref{bernfig11}  shows the results of  Shukurov et al. (2006) for a model of the Galactic dynamo.  The equations shown represent the generalization of Eqs. (\ref{9}) and (\ref{10}) to the $\alpha-\Omega$ dynamo  with only vertical $z$ derivatives retained  and
with  an advective flux term that ejects small scale magnetic helicity from the system. The figure illustrates that the large scale field decays when this flux  is too small, illustrating its important role in sustaining the EMF.
In the solar context,  a similar circumstance arises: K\"apyl\"a et al. (2008) found from simulations that an LSD is produced by convection + shear when surfaces of constant shear were aligned toward open boundaries allowing a helicity flux whereas Tobias et al. (2008) found no LSD when shear was aligned toward periodic boundaries  disallowing helicity fluxes. 

The role and potential observability of such global helicity fluxes  are  exemplified schematically in Fig. \ref{bernfig12},
originally in the context of the sun (Blackman \& Brandenburg 2003).
The figure  shows the same principles as the comparison between Figs. \ref{bernfig5} and \ref{bernfig7} but with shear and buoyancy. In the northern hemisphere, the field exhibits a right handed writhe and a left handed  small scale twist as the structure buoyantly emerges into the corona.  The emergent separation of scales of magnetic helicity was verified by simulation of a buoyant, writhed tube. 
Such  structure might in fact be evident in the TRACE image of  coronal loop shown in Fig. \ref{bernfig13} from Gibson et al. (2002).  The figure shows left hand twist and right handed writhe.  A key point is that such  ejections could be an essential
 part of sustaining the fast solar cycle, not just an independent consequences of magnetic field generation.
 
  This segregation  of helicity signs is consistent with  other evidence that the northern hemisphere exhibits primarily small scale left handed twist and larger scale right handed writhe, with these reversed in  southern hemisphere
   (Rust and Kumar 1996; Pevtsov et al. 2008, Zhang et al. 2010; Hau \& Zhang 2011). These features seem to be   invariant with respect to solar cycle,  as would be predicted by helical dynamos, even when the field itself reverses sign.  Note also  that   $H\alpha$ filaments seem to exhibit  ÒdextralÓ (right handed) twist in North and ÒsinstralÓ (left handed) in south (Martin and McAllister 1997) BUT: right handed $H\alpha$  filaments may be supported by left handed fields and vice versa (Rust 1999)

There have been efforts to measure the rate of injection of magnetic helicity (Chae 2004; Schuck 2005; Lim et al. 2007), particularly, its gauge invariant cousin: the ``relative magnetic helicity" (Berger \& Field 1984;  the  difference between actual magnetic helicity and that of a potential field)  by tracking footpoint motions. It can be shown that the footprint motions
provide a direct measure of this input rate (D\'emoulin \& Berger 2003).
There have also been efforts to relate  the current helicity to the injection rate of magnetic helicity (Zhang et al. 2012).
Ultimately, measuring the detailed spectra of current helicity and relative magnetic helicity injection into the corona and solar wind (Brandenburg et al. 2011) is a highly  desirable 
enterprise  for the future.
Commonly, observational work has focused on the component of twist with the current along the line of sight,  but measuring the full current and magnetic helicities require all three components of the field.

\subsection{Helicity Fluxes in Accretion Disks and Shearing Boxes}

Essentially all  MRI unstable simulations with large enough vertical domains,
whether stratified, unstratified,  local or global, show the generation of large scale toroidal fields
of the same flux for  cycle periods of $\sim 10$ orbits 
(Brandenburg et al. 1995, Lesur \& Ogilvie 2008, Davis et al. 2010; Simon et al.  2011; Guan \& Gammie 2011; Sorathia et al. 2012; Suzuki and Inusuka 2013;
Ebrahimi \& Bhattacharjee 2014). The patterns indicate a large scale dynamo operating contemporaneously with the small scale dynamo.
The key unifying property of all of these cases  is  mean field aligned EMF, that is $\lb \emfb\cdot \bbB\rb$.
The explicit form  of $\emfb$ and the terms that contribute to it  can depend  on  the boundary and stratification conditions and on the particular procedure for large scale averaging, but there is considerable similarity among the dynamos operating in these simulations.   
Sorting out  whether and which flux terms are important for  specific averaging and initial conditions  is an ongoing task.
To see the issues at hand,  I highlight  approaches to  LSD  modeling  in shearing boxes without  using helicity fluxes and then compare to those that do.  

 The left panel of  Fig. 16,  adapted  from Simon et al. (2011),   shows the toroidal magnetic field from a  shearing box MRI simulation. The simulation used vertical stratification and  periodic boundaries in the $x$ and $y$ (radial and azimuthal) and outflow  boundaries in $z$ (vertical) directions. The toroidal field was calculated by averaging over $x$ and $y$ and over $z\le 5H$ where $H$ is a density scale height. There was an initial net toroidal field in the box, but the left panel shows that the toroidal field reverses every 10 orbits in the saturated state.  The right panel shows  the use of  model equations from an  $\alpha-\Omega$ dynamo that Simon et al. (2011) adopted from Guan \& Gammie (2011)  tuned to match the simulation.  The equations  have only a shear term and a  loss term from buoyancy in the torioidal field equation. The radial field equation has a buoyancy loss term and the $\alpha=\alpha_2$ dynamo term. 
There are no  helicity dynamics in this empirical set of equations; those dynamics would at most y be  hidden in the empirically determined $\alpha_2$.  Nevertheless,   this quasi-empirical set of equations does well to model the field seen in the simulations.

In this example of Simon et al. (2011),  like those of the first   analyses of cycle periods in shearing box simulations 
 (Brandenburg et al. 1995; Brandenburg \& Donner 1997), the sign of the  $\alpha=\alpha_2$ coefficient is found to be inconsistent with the standard  20th century textbook kinetic helicity. Additional features such as  density stratification and rotation lead to higher order turbulent anisotropy and inhomogeneity that change the dominant sign of $\alpha$  appropriately  \citet{1993A&A...269..581R},
 and is consistent with the role of magnetic buoyancy  (Brandenburg 1998).
In this context, Gressel (2010)  further  looked at the behavior of the dynamo $\alpha$ coefficient in stratified MRI shearing box (with outflow vertical boundaries) simulations
and found that the sign of the  dynamo $\alpha$ coefficient was  consistent with the generalizations of  \citet{1993A&A...269..581R}, and also consistent with the sign of the current helicity  correction term to $\alpha$ in Eq. (\ref{7aab}), as if the current helicity contribution may be sourced by magnetic buoyancy.   Gressel's  results are shown in Fig. \ref{bernfig15} and his averaging was taken over radius and azimuth.


The aforementioned approaches to  MRI  LSDs do not involve
helicity fluxes, but calculations of EMF sustaining fluxes in this context have been emerging
 (e.g Vishniac \& Cho 2001; Vishniac 2009; K\"apyl\"a \& Korpi 2011; Ebrahimi \& Bhattacharjee 2014)
 (See also Vishniac \& Shapovalov 2014 for an isotropically forced case with shear). 
In particular, Ebrahimi \& Bhattacharjee (2014)   studied an MRI unstable cylinder with conducting
boundaries. They wrote down mean quantities averaged over azimuthal and vertical directions,  leaving the radial direction unaveraged. They  looked explicitly at the terms in the helicity conservation equation  Eq. (\ref{5aa}) and found that the electromotive force is well matched by the local flux terms  measured from  direct numerical simulations. Their results are shown in  Fig. \ref{bernfig16}.

To  calculate the specific form of these fluxes by brute force in  a   mean field theory involves 
expanding the fluctuating quantities in terms of mean field quantities via the dynamical equations for those fluctuating quantities and 
a closure (e.g. Brandenburg and Subramanian 2005). Such efforts are ongoing.  One such flux that emerges from such a procedure that of Vishniac  \& Cho (2001). Ebrahmi \& Bhattacharjee (2014) also plot this latter  flux as seen in the let panel of Fig. \ref{bernfig16}. They find that it is much smaller than the total flux term directly calculated from the simulations, and thus is subdominant. Recall that Shukurov et al. (2006) in Figure \ref{bernfig11} considered an advective flux term that is another candidate flux term that emerges
from mean field theory.  Sur et al. (2007) also assessed the role of the Vishniac-Cho flux semi-analyticially in the galactic context and found it can be helpful if the mean magnetic field is above a threshold value to begin with.
Vishniac \& Shapovalov (2014)   considered an isotropically forced periodic box with linear shear,
 (without the coriolis force) and  find that the  Vishniac-Cho 
flux is dominant.   Their averaging procedure involves averaging over the entire box and filtering by wavenumber to distinguish contributions from mean and fluctuating components.  This procedure is different from that of Ebrahimi \& Bhattacharjee (2014) discussed above.  See also Hubbard  \& Brandenburg (2011) however, who suggest that the choice of gauge  influences whether
the Vishniac-Cho flux is important in numerical simulations.

 In general,  the different circumstances and averaging procedures of mean field quantities between 
simulations of e.g. Simon et al. (2011), Ebrahimi \& Bhattacharjee (2014) and those of Vishniac \& Shapovalov (2014) highlight the need for clarity in tailoring the specific mean field theory  to capture the dominant contributions  for different combinations of forcing, boundary conditions, and averaging procedures.  There is also an opportunity  to combine mean field theories for local transport (e.g. Ogilvie 2003; Pessah et al. 2006) with those of mean field dynamos and large scale transport  to fully model angular momentum transport in  accretion disks.

\section{Gauge Issues}

A subtle aspect of the helicity density equations (\ref{5aa}) and (\ref{6aa}) is the 
issue of gauge non-invariance.  In the absence of boundary flux terms or time dependent terms,   the  magnetic helicity is gauge invariant relative to an arbitrary choice of  initial value. This is a straightforward consequence of Eq. (\ref{6bf}).  Since $\bfE\cdot\bfB$ is gauge invariant under any circumstance,
  the sum of the time dependent term on the left minus the divergence term on the right is always gauge invariant (even though
the flux itself need not itself be gauge invariant).   
If the flux term vanishes, then the the time derivative term is gauge invariant.
  If the time derivative term vanishes, then the leftover flux term is gauge invariant. This  was used by Blackman \& Field (2000b)
  to estimate an energy associated with the ejected magnetic helicity into coronae by a steady state dynamo.
  Mitra et al. (2010) show numerically that indeed the diffusive magnetic helicity fluxes that arise naturally across the mid plane in a system forced 
  with oppositely signed kinetic helicities are gauge invariant. In the  steady-state the divergence is invariant at every point so one can obtain the spatial dependence of the flux. These principles were also verified in Hubbard \& Brandenburg (2010) and  apply even to oscillatory dynamos by first identifying a gauge for which the helicity is steady (Del Sordo et al. 2013),  which eliminates the dependent terms and the divergence term emerges as gauge invariant.
  
Additional subtleties of gauge invariance in the different  context of shearing boxes are discussed in Canderlaresi et al. (2011) and Hubbard \& Brandenburg (2011).

Gauge non-invariance is closely related  to
  the fact that for open boundaries, the amount of external field linkage is not in general specified.
That is,  the amount of field linkage inside the boundary can be the same for different amounts of exterior linkage. Fixing the gauge for the vector potential removes this ambiguity in that gauge.  In actuality,    one can  choose a gauge, work in this gauge to study helicity dynamics, and then  calculate physical quantities that are gauge invariant.   The gauge non-invariance does not change the role of magnetic helicity as an intermediate conceptual tool. 

However to interpret physically the magnetic helicity, gauge invariant versions can be helpful. 
Subramanian \& Brandenburg (2006)  developed a  generalized local helicity density whose evolution reduces to the above Eqs. (\ref{5aa}) and (\ref{6aa})  in the absence of flux terms. In the presence of flux terms, their equation has the same form but with a   different helicity density that is gauge invariant. For a turbulent system, with large scale separation between fluctuating and mean quantities, 
their gauge invariant helicity density associated with small scale quantities is similar to what is obtained 
for the usual magnetic helicity density in  the Coulomb gauge 
because the gauge variant boundary terms become small in their  averaging procedure.
In the context of the Galaxy, Shukurov et al. (2006)  solved the mean field induction equation for $\bbB$ using this  
gauge invariant helicity density.

The  gauge invariant  relative magnetic helicity  (Berger \& Field 1984; Finn \& Antonsen 1985; Bellan 2000) referred to earlier,  was developed for more direct  interpretations of observations originally in the solar context.
 This quantity involves separating global  space into two parts,
a  region of physical interest and the exterior to this region. The relative magnetic helicity is specifically the difference between the magnetic helicity of the system integrated over global  space   minus that associated with an integral over global space where the field in region of physical interest is replaced  by  a  vacuum, potential field, namely $H_R\equiv  \int \bfA\cdot \bfB dV -\int \bfA_p\cdot \bfB_p dV$, where 
$\bfB_p=\curl \bfA_p$ and $\bfB_p$ is the potential field. 
In this way the external linkage is removed, and   what remains is gauge invariant. 
The time evolution equation  for $H_R$ is then (Berger \& Field 1984) 
\beq
\partial_t H_R =  -2c\int \bfE\cdot \bfB dV  -\int\nabla \cdot( \bfE\times \bfA_p)dV.
\eeq
All terms in this generalized relative magnetic helicity conservation equation are gauge invariant.
 Not only is the divergence term gauge invariant  but the flux itself is gauge invariant.
D\'emoulin \& Berger (2003) have shown how the rate of injection of relative magnetic helicity into the solar corona
depends on measurable  quantities  at the footprints of the anchoring fields.  Sorting out the relation between the relative magnetic helicity,  the current helicity, and the gauge variant magnetic helicity in theoretical calculations  warrants  further attention.

\section{Summary and Conclusions} 

Tracking magnetic helicity helicity in MHD systems  is  an important, unifying tool to understand the 
the processes by which large scale magnetic fields form and evolve in both astrophysical and laboratory plasmas.
The purpose of this review has been to provide one  path through the forrest as a conceptual primer to the 
literature.  Three key principles underlie the role of magnetic helicity in all contexts: (1) magnetic 
helicity is is a measure of twist, linkage or writhe; (2) magnetic helicity is better conserved than magnetic energy under most circumstances for a closed system with or without velocity flows; (3) the energy in a helical magnetic field is minimized when the field relaxes to the largest scale available, consistent with the boundary conditions.

The importance of magnetic helicity for large scale field generation is evident from early studies of helical MHD turbulence (Pouquet et al. 1976), but incorporating its role into a 21st century dynamical mean field theory has emerged only in the pas decade or so.
The 20th century textbook dynamos, unlike 21st century theory, do not conserve magnetic helicity and as such were unable to 
predict or reveal how LSDs saturate.  A key aspect of 21st century theory is that the growth of large scale fields, facilitated by an EMF,
involves growth of large scale magnetic helicity and small scale magnetic helicity of the opposite sign. In the absence of helicity fluxes,  the
small scale build up suppresses further growth of large scale field. In the case of sheared rotators,  unless helicity fluxes can remove the offending small scale  magnetic helicity, the large scale field  not only saturates  but  may even decay. Alternatively expressed, 
it seems that astrophysical dynamos, like laboratory plasma dynamos may involve an  EMF that is commonly aided or sustained by 
the divergence of small scale helicity fluxes.   In a quasi-steady state for the sun, a crude minimalist prediction is that
both signs of helical magnetic fields  should appear in the northern hemisphere with small scale left handed structures and right handed large scale structures with the reversed combination in the southern hemisphere. More efforts to measure the spectral
distribution of helical fields in the solar corona and wind would be valuable. 
Large scale dynamo models for the galaxy and for  accretion disks that incorporate magnetic helicity dynamics have also been emerging.

Accretion disks pose an interestingly rich opportunity of study for helicity dynamics and large scale dynamos because traditionally large scale dynamo theory has been studied independent of theories of angular momentum transport. The ubiquity of large scale dynamos seen in simulations and the ubiquity of observed astrophysical coronae and jets 
 indicates that a significant contribution to angular momentum transport comes from large scale fields. This needs to be incorporated into a combined mean field accretion disk theory that captures local and large scale angular momentum transport, and large scale field growth.   

Coronae of stars, disks, and laboratory plasmas are all magnetically energy dominated. In magnetically dominated environments   the principles of magnetic helicity evolution have long been helpful to understand the evolution of magnetic structures subject to their foot-point motions. The helicity injection by  foot-point motions is analogous to  injection of small scale helicity in laboratory devices,
where the  system responds by relaxing the helicity to large scales  In astrophysical coronae, it is likely that some contribution from both signs rather than a single sign are injected, so the relaxation process must take this into consideration globally,even if local structures are injected with primarily one sign.  It was in fact in the context of  laboratory plasma magnetic relaxation where the importance of helicity fluxes was first identified.

Finally, as reviewed herein, the basic properties of magnetic helicity also underlie its role in making large scale helical fields resilient to turbulent diffusion.  Recent work on this topic for closed systems may strengthen the potential efficacy of fossil field origin of large scale fields in some astrophysical contexts. A helical field (in the absence of global helicity fluxes) is much more resilient to turbulent diffusion than  non-helical large scale fields. The effect is best understood not as reduction of the turbulent diffusion coefficient, but rather as a competition between the unfettered turbulent diffusion and additional competing  tendency for helicity to relax back toward the largest scales. The  driver for this inverse transfer is the very small scale magnetic helicity that is sourced by initial diffusion from the large scale helicity in the first place.

\section*{Acknowledgements}
I acknowledge support from NSF grant  AST-1109285, and thank the organizers of the ISSI Workshop on "Multi-Scale Structure Formation and Dynamics of Cosmic Plasmas",  and  the  organizers of the Lyman Spitzer 100th birthday conference  for engaging meetings in  Bern and Princeton respectively.  I also acknowledge particular  discussions with  P. Bhat, A. Bhattacharjee, A. Brandenburg, F. Ebrahimi, G. Field, A. Hubbard,  J. Stone, F. Nauman,  and K. Subramanian.

\vfill
\eject

\begin{figure}
  \includegraphics[width=\columnwidth,trim=0 0 0 0]{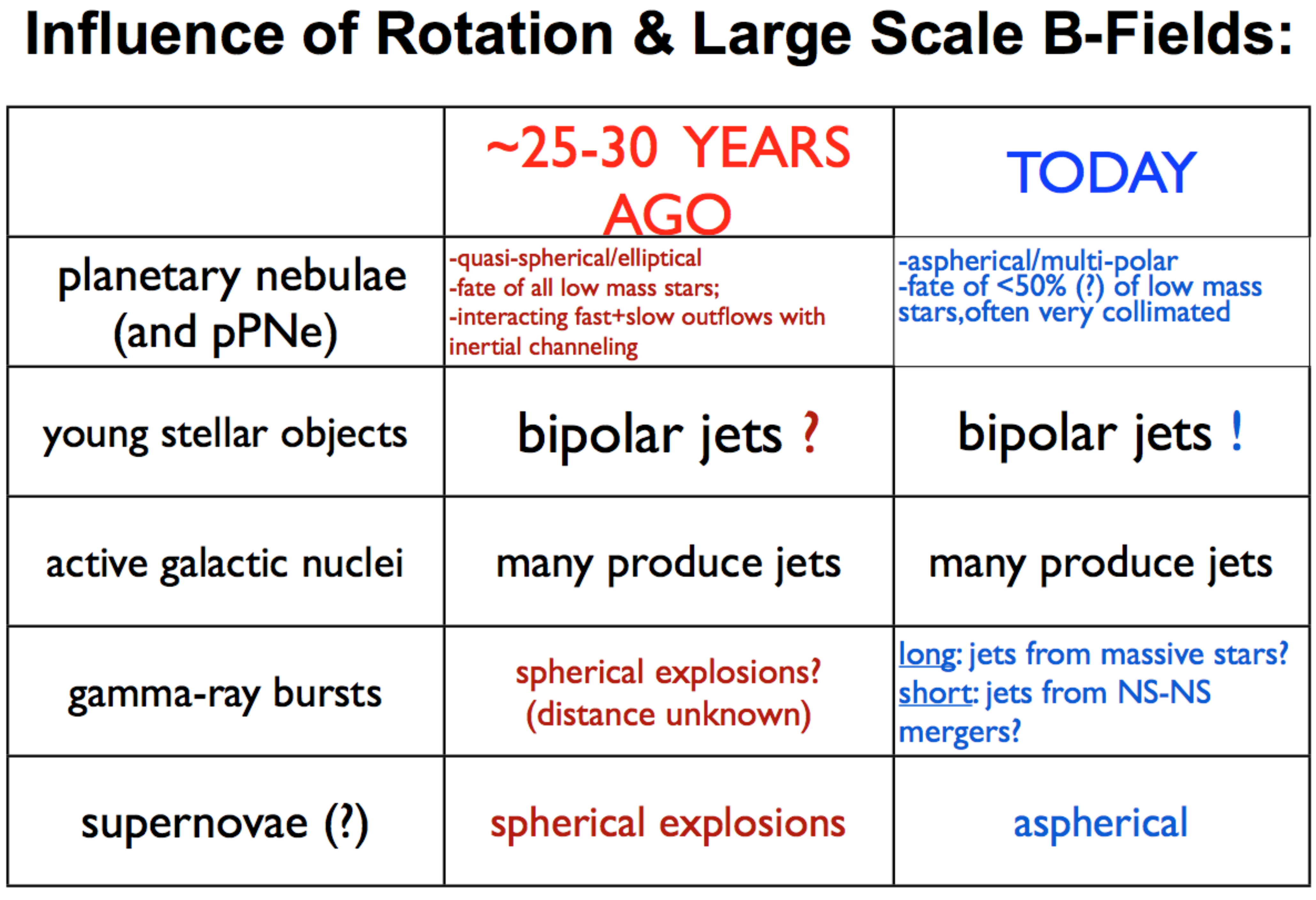}
  \caption{Chart highlighting an emerging realization in astrophysics that more sources originally perceived as spherical may in fact be bipolar, and thus harborers of large scale magnetic field mediated jets. }
  \label{bernfig1}
\end{figure}

\begin{figure}
  \includegraphics[width=\columnwidth,trim=0 0 0 0]{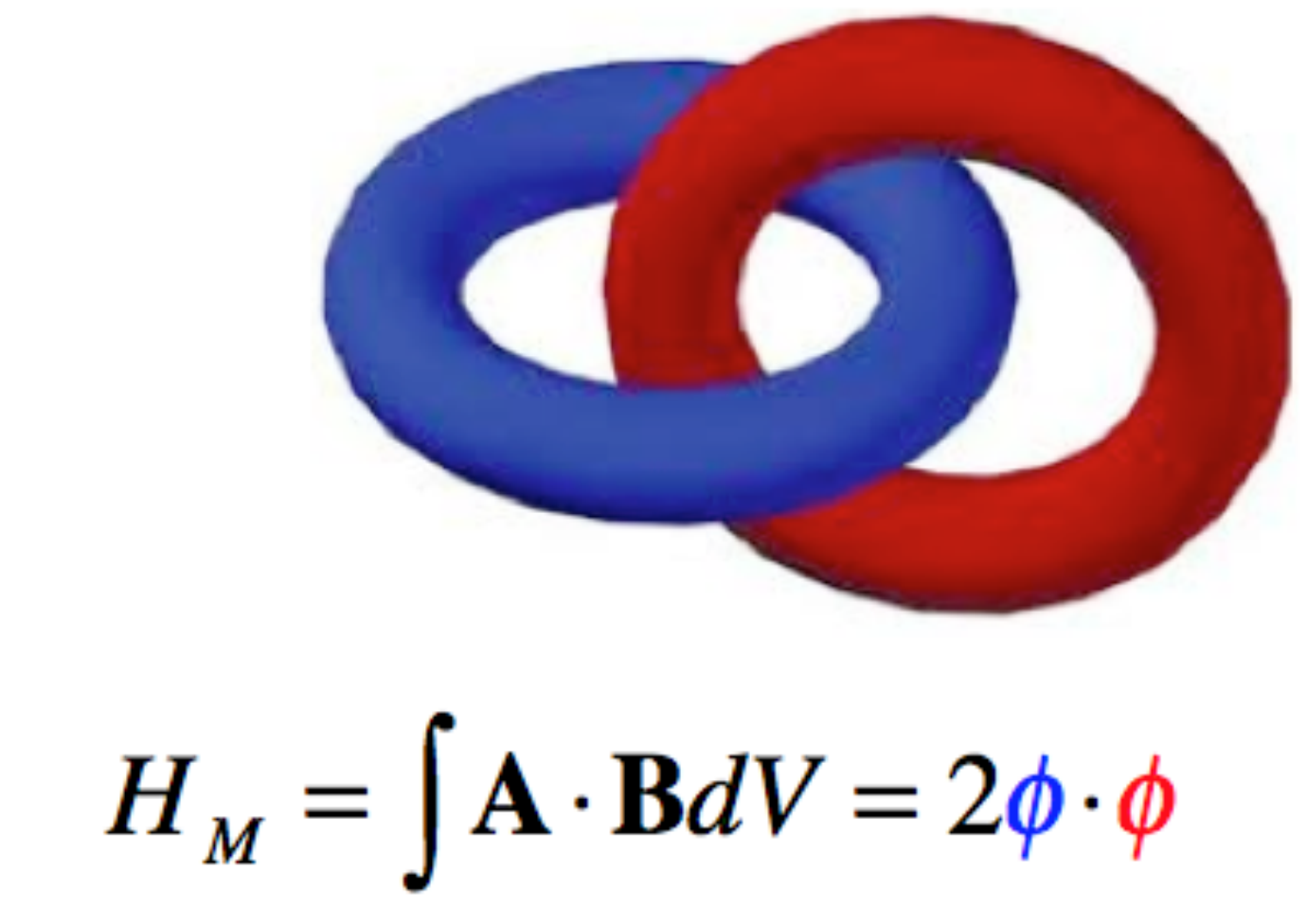}
  \caption{Magnetic helicity of two linked flux equals twice the product of their magnetic fluxes. }
  \label{bernfig2}
\end{figure}

\begin{figure}
  \includegraphics[width=\columnwidth,trim=0 0 0 0]{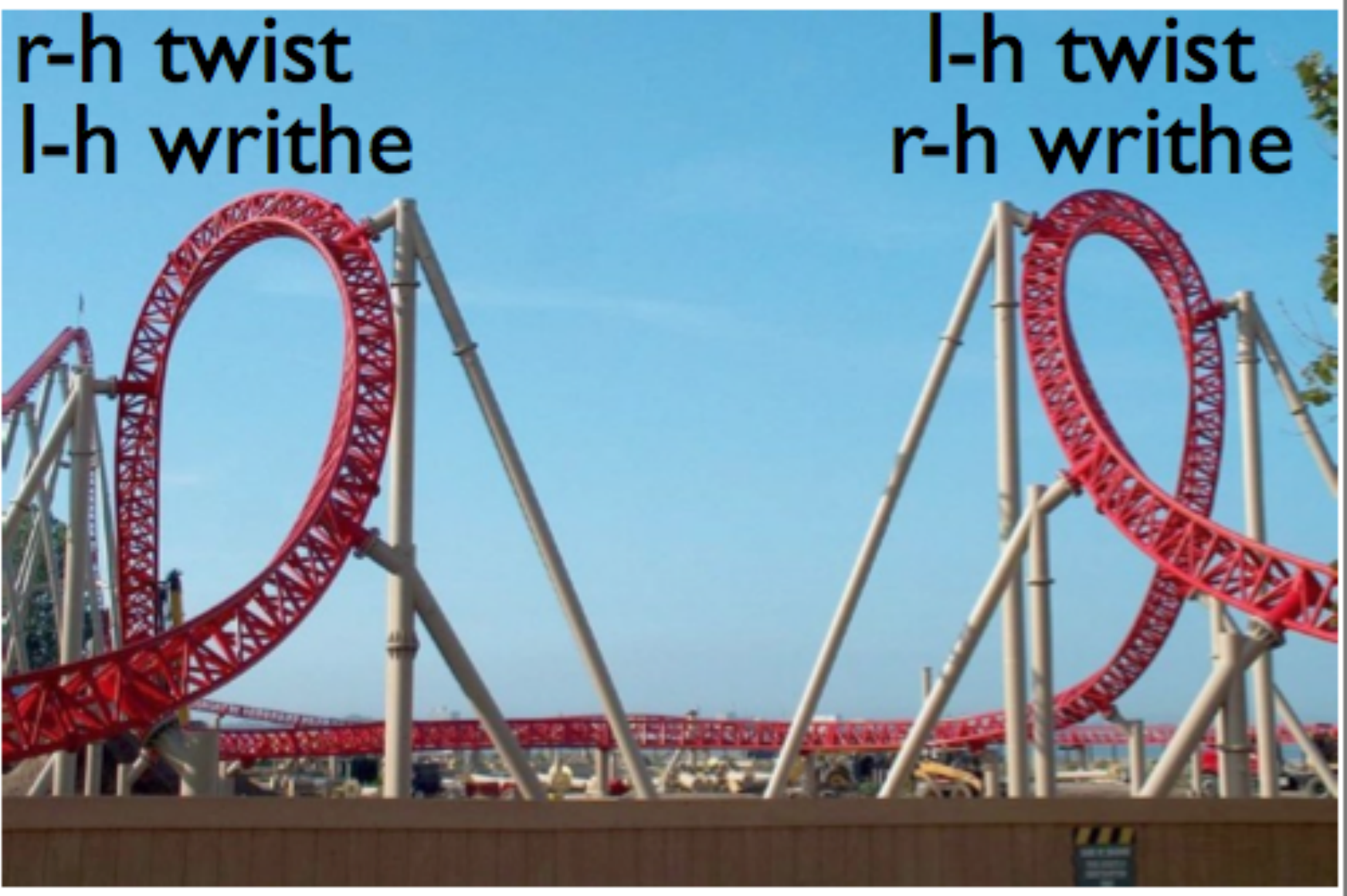}
  \caption{Picture of a "twisted horseshoe roll" local element  of a roller coaster at Cedar Point in  Sandusky, Ohio, USA. The large scale loops are "writhe" as defined in the text and the rotation of the tracks along the direction of the coaster path is "twist" . The loop on the left has left-handed writhe and right handed twist. The loop on the right has right-handed writhe and left-handed twist. The overall trip on the coaster is a closed path, the rest of which is not shown.}
  \label{bernfig3}
\end{figure}

\begin{figure}
  \includegraphics[width=\columnwidth,trim=0 0 0 0]{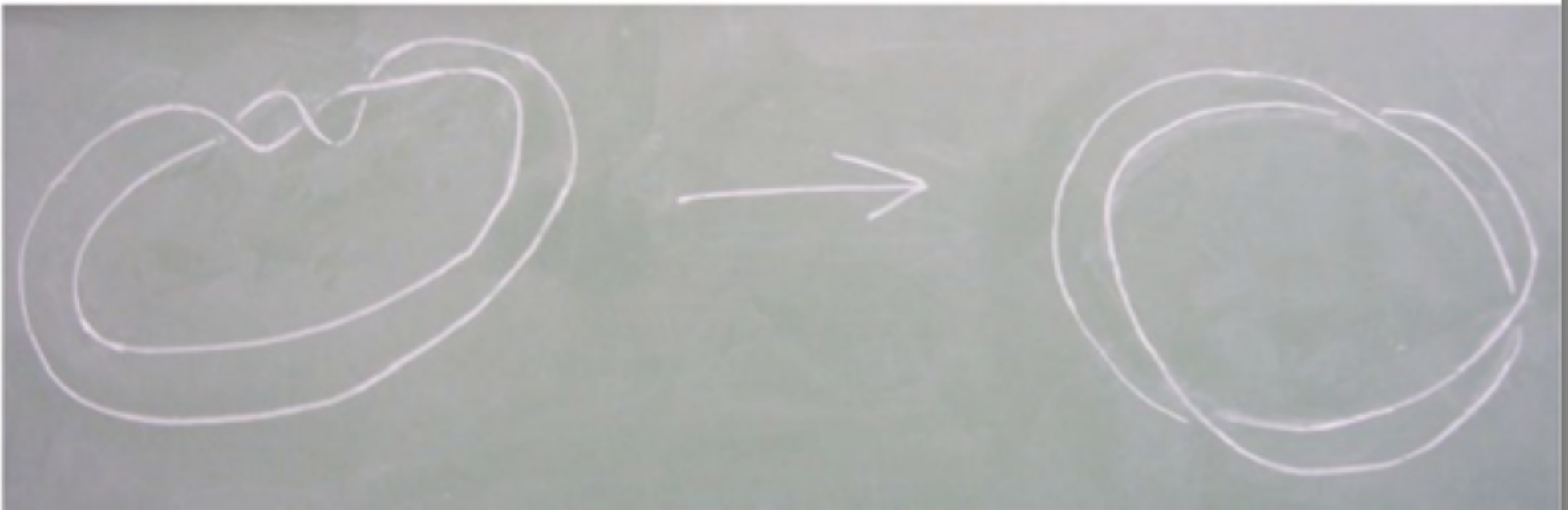}
  \caption{The sketch illustrates two concepts. The first is  magnetic relaxation: The small scale twist of the ribbon on the left  has relaxed to the largest scale available to it in the right figure. The right configuration is a lower energy state fire the same magnetic helicity as that in the left.
  The second concept is that  these structures can also be viewed  as two strands,  so that one can qualitatively appreciate some equivalence between twist of a ribbon and linkage of a pair of strands. In the text, this is made more quantitative for a different example shown in Fig. 5.}
  \label{bernfig4} 
\end{figure}

\begin{figure}
\minipage{.42\textwidth}
  \includegraphics[width=\linewidth]{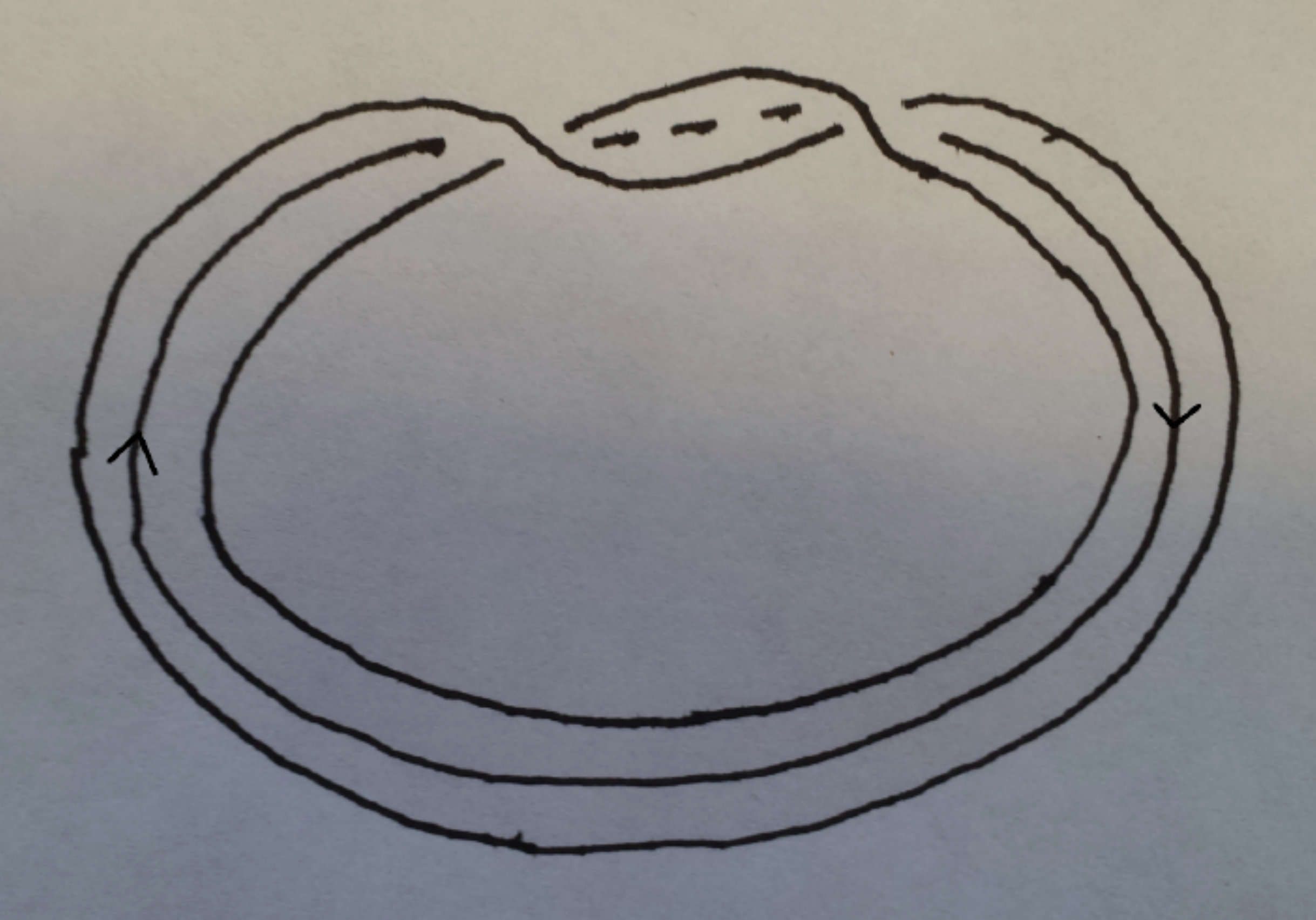}
\endminipage\hfill
\minipage{.6\textwidth}
  \includegraphics[width=\linewidth]{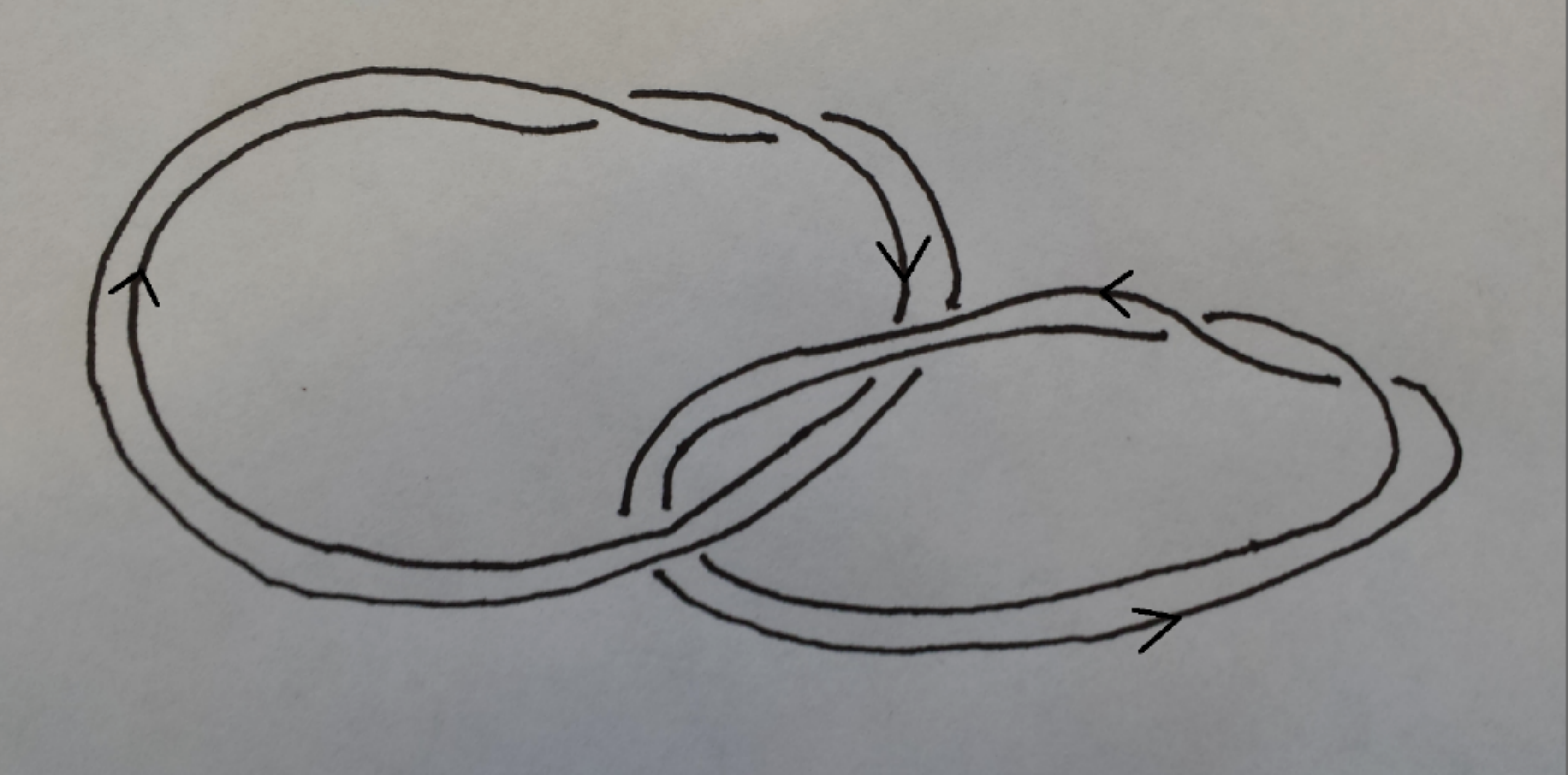}
\endminipage\hfill
\caption{The two panels show schematics of magnetic ribbon systems with equivalent amounts of magnetic helicity. The left panel shows a single twisted ribbon. If this were represented by a strip of paper, cutting along the centerline then results in the right panel. The single thick ribbon with one right handed twist has been converted to two linked ribbons of 1/2 half the width of the original, and each with a right handed twist. This figure illustrates why helicity can be represented as twist, linkage, or some combination of the two. }
  \label{bernfig4.2}
\end{figure}

\begin{figure}
\minipage{.555\textwidth}
  \includegraphics[width=\linewidth]{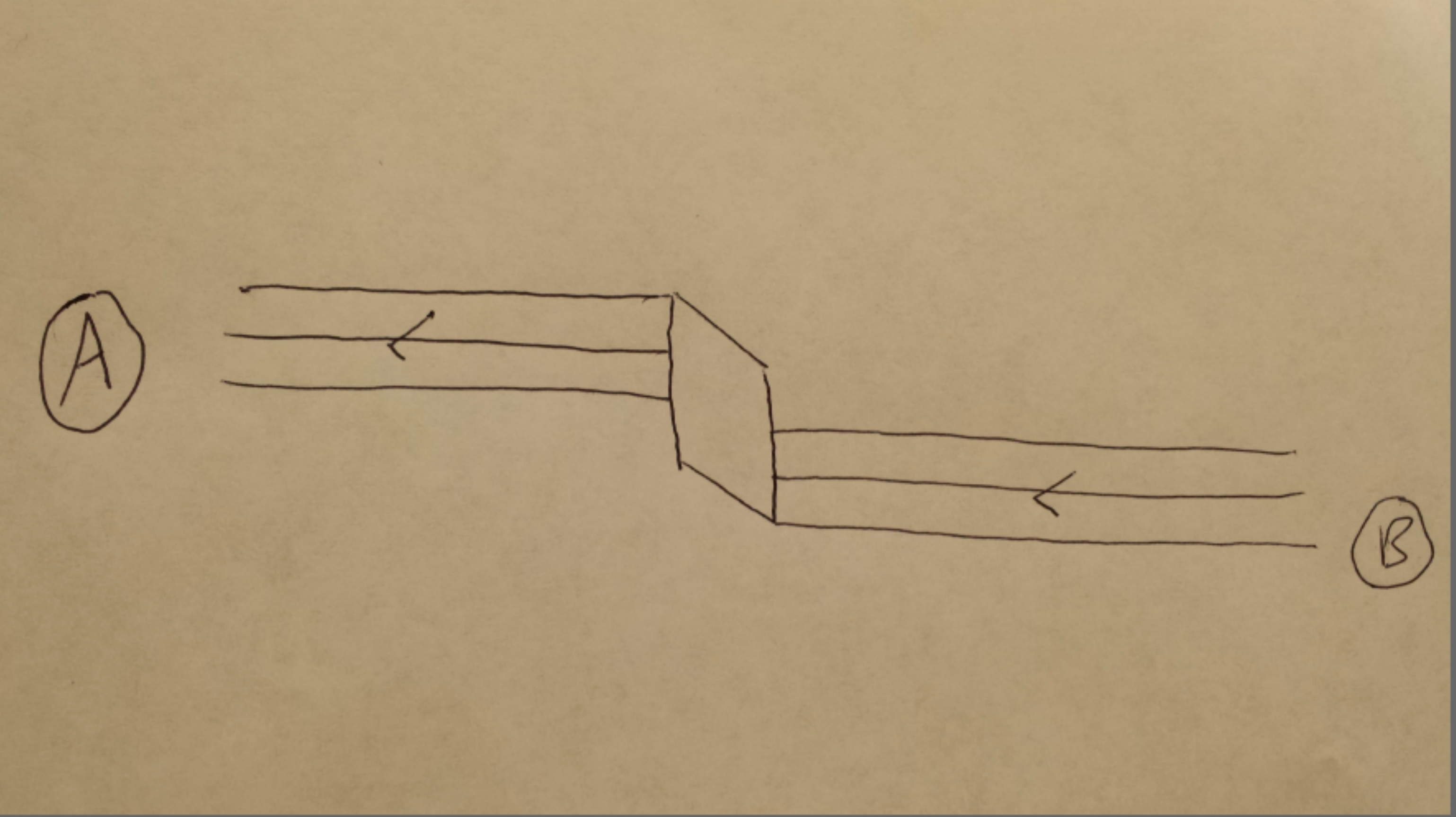}
\endminipage\hfill
\minipage{.56\textwidth}
  \includegraphics[width=\linewidth]{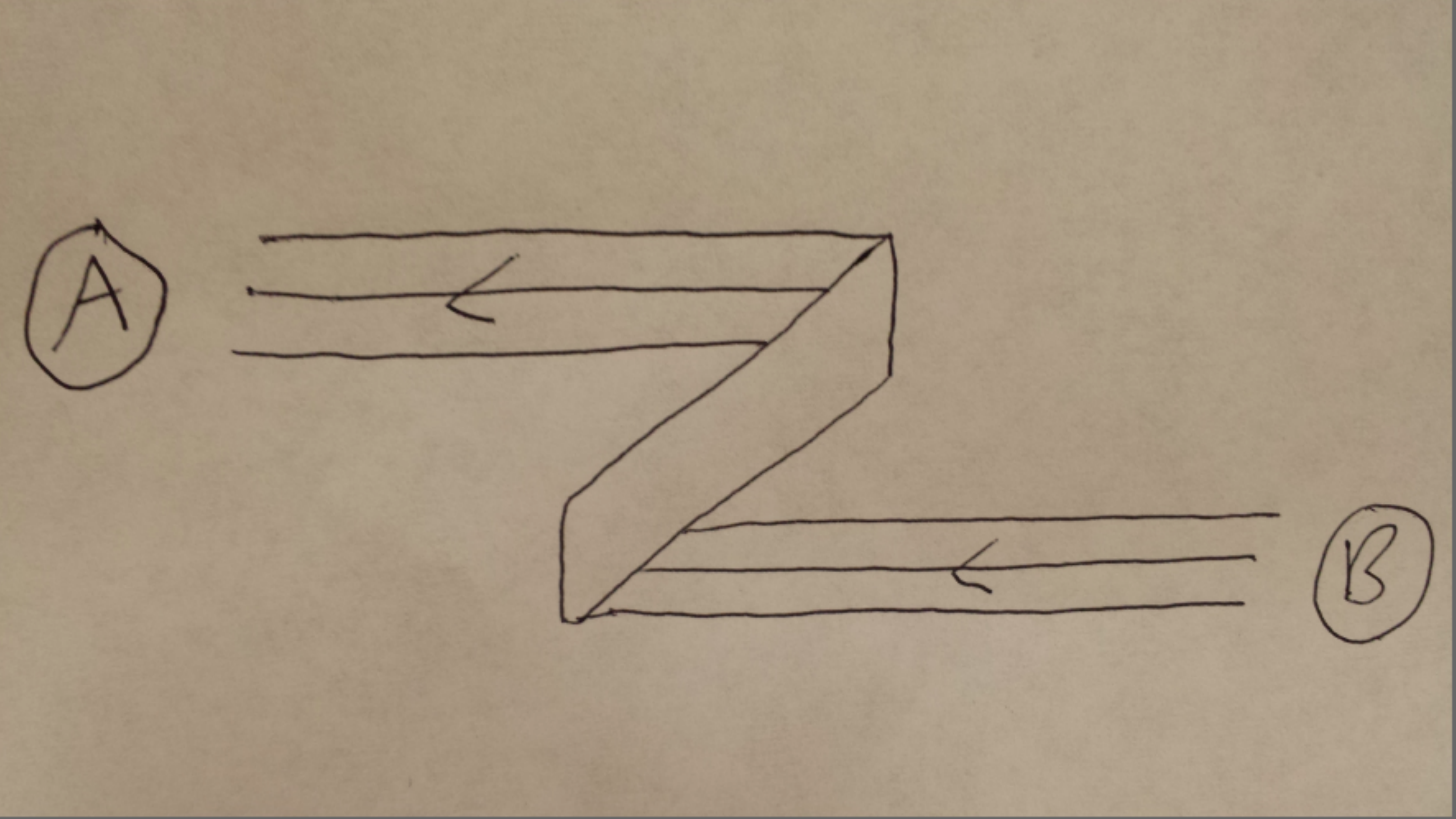}
\endminipage\hfill
\minipage{.6\textwidth}
  \includegraphics[width=\linewidth]{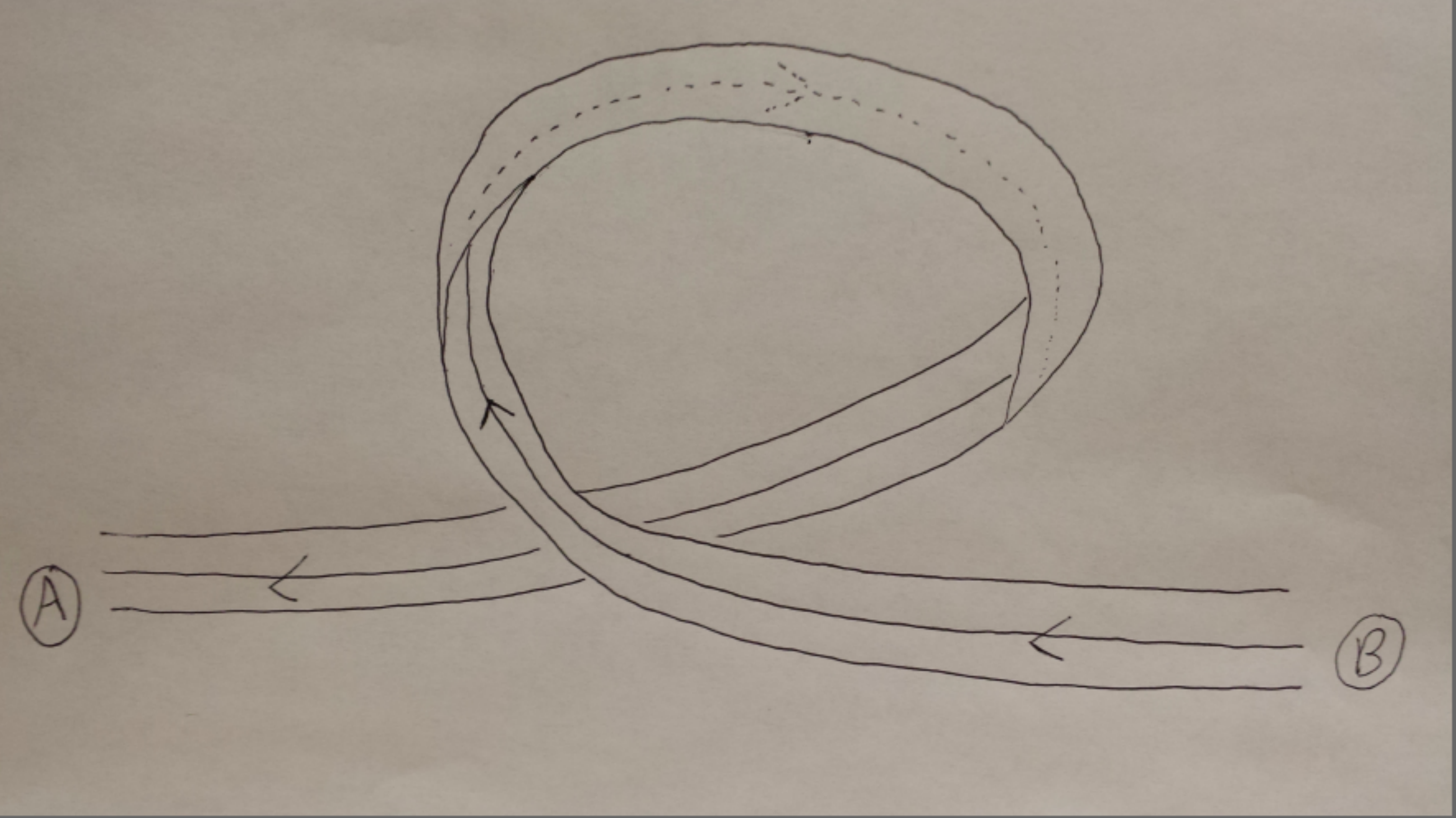}
\endminipage
\caption{The three panels show how to transform twist helicity into writhe helicity. (top left) Start with a straight flat ribbon with one unit of right handed twist.
One could imagine that the ends A and B are identified with each other and so it is a mathematically a closed ribbon loop with a twist.
(top right) Push the ends A and B toward each other until the ribbon buckles upward.  (bottom) A  side view  shows the writhe. The result is a writhed, closed loop.}
  \label{bernfig4.3}
\end{figure}

\begin{figure}
\minipage{.5\textwidth}
  \includegraphics[width=\linewidth]{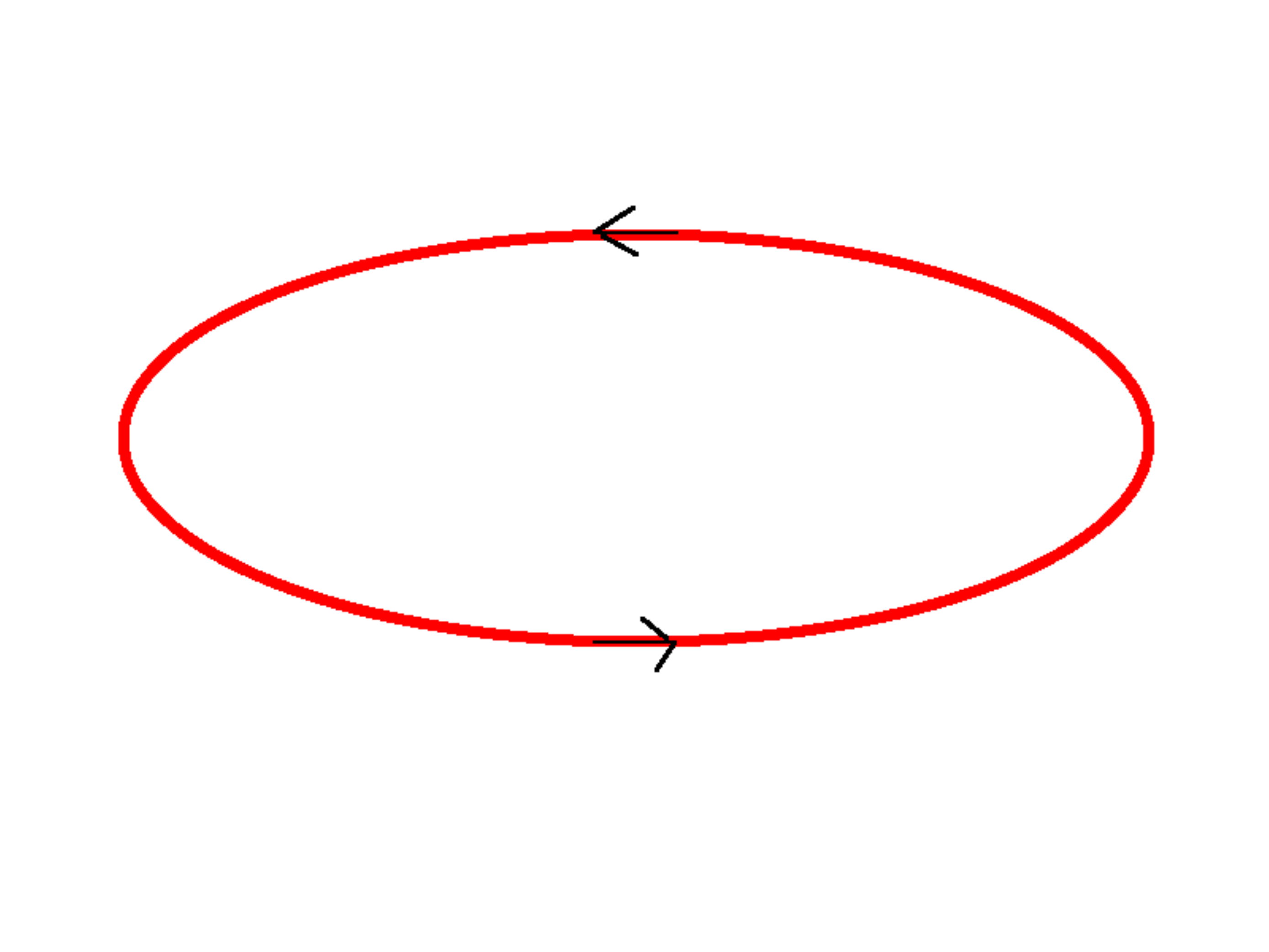}
\endminipage\hfill
\minipage{.5\textwidth}
  \includegraphics[width=\linewidth]{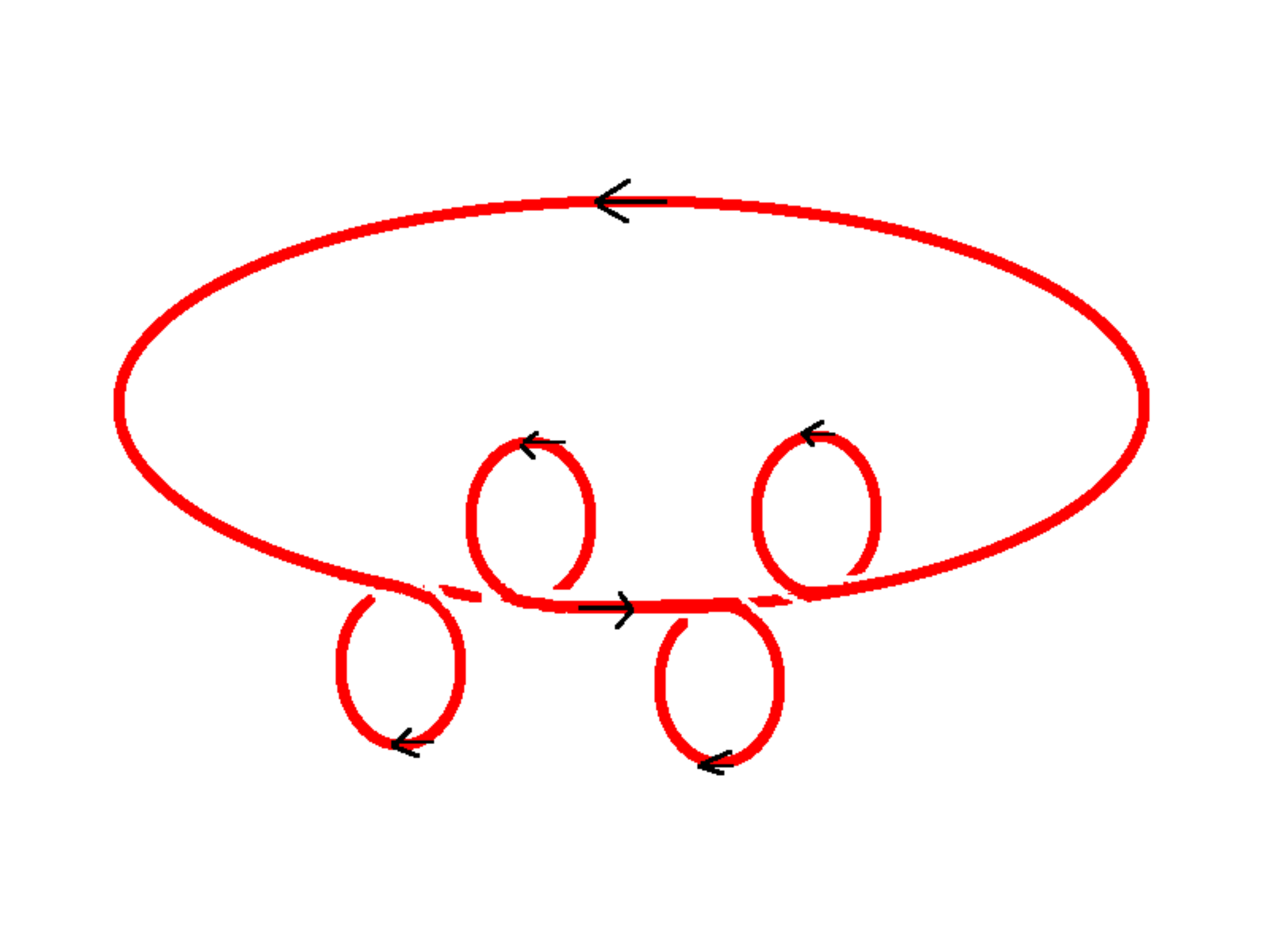}
\endminipage\hfill
\minipage{1\textwidth}%
  \includegraphics[width=\linewidth]{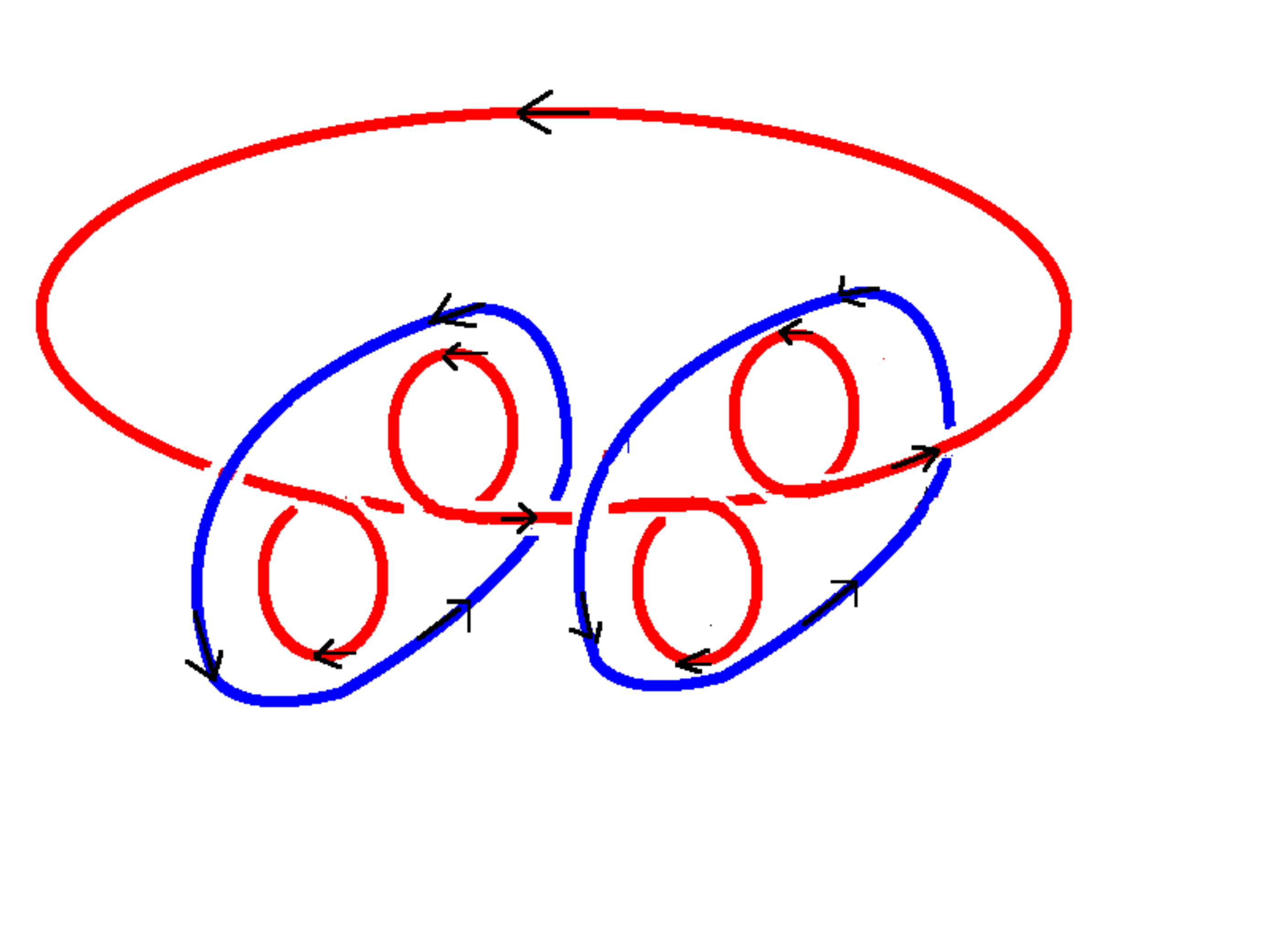}
\endminipage
\caption{Two-stage schematic for  magnetic field structure in an  $\alpha^2$ dynamo driven by negative  kinetic helicity in the conventional 20th century approach with the magnetic field represented as lines.
  The top left panel shows a large untwisted  toroidal magnetic field ribbon.
The top right panel shows the action of  kinetic helicity on the initial ribbon. The small scale negative kinetic helicity produces each  the four small scale poloidal loops.  Each small loop  incurs a writhe (or overlap) of positive (=right-handed) magnetic helicity.    The two intermediate scale poloidal loops  encircling the small scale loops in the bottom panel represent the resultant mean poloidal field 
averaged   separately over each pair of  loops.  These intermediate scale  loops are  linked to the initial large scale torioidal loop.  Since a single linked pair of ribbons has  2 units of  magnetic helicity, we see that the two poloidal loops linking the torioidal loop have a total 4 total units of right handed magnetic helicity. This helicity in the "large scale field" has come from zero initial magnetic helicity
and thus cannot be correct for MHD at large $R_M$ which conserves magnetic helicity. The diagram does not 
not account for the missing small scale magnetic helicity of opposite sign. (Compare to Fig. 8)}
  \label{bernfig5}
\end{figure}


\begin{figure}
\minipage{.5\textwidth}
  \includegraphics[width=\linewidth]{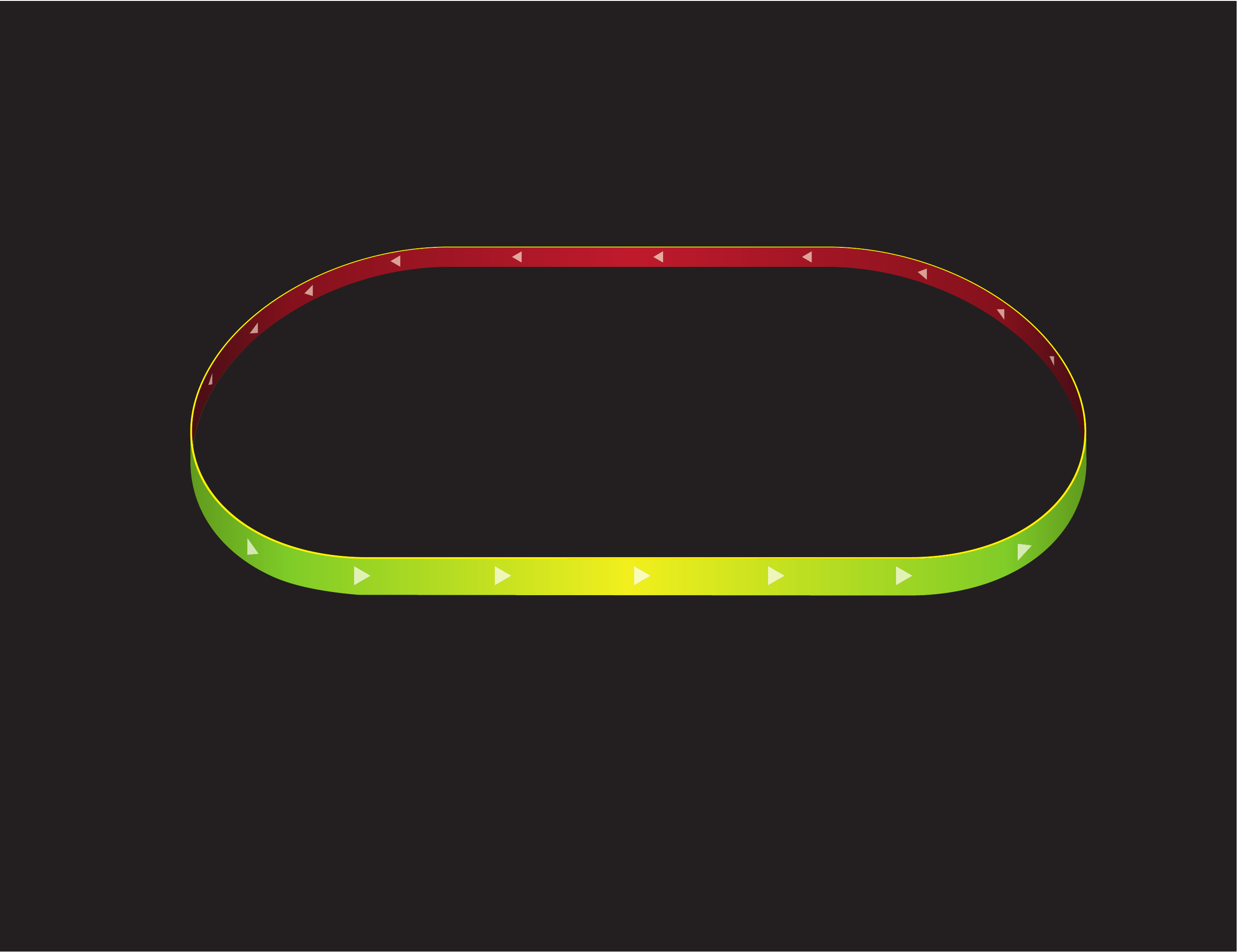}
\endminipage\hfill
\minipage{.5\textwidth}
  \includegraphics[width=\linewidth]{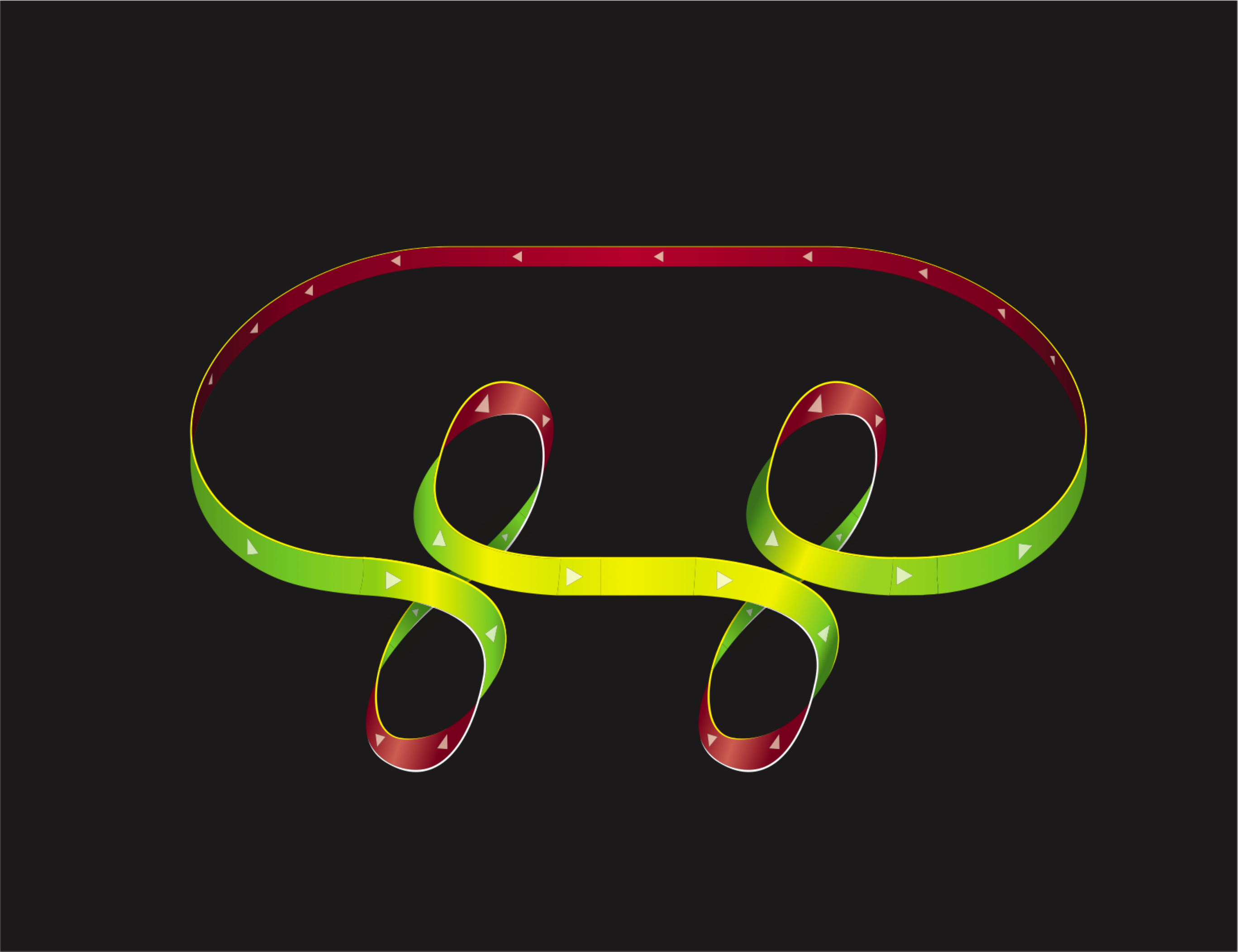}
\endminipage\hfill
\minipage{1\textwidth}%
  \includegraphics[width=\linewidth]{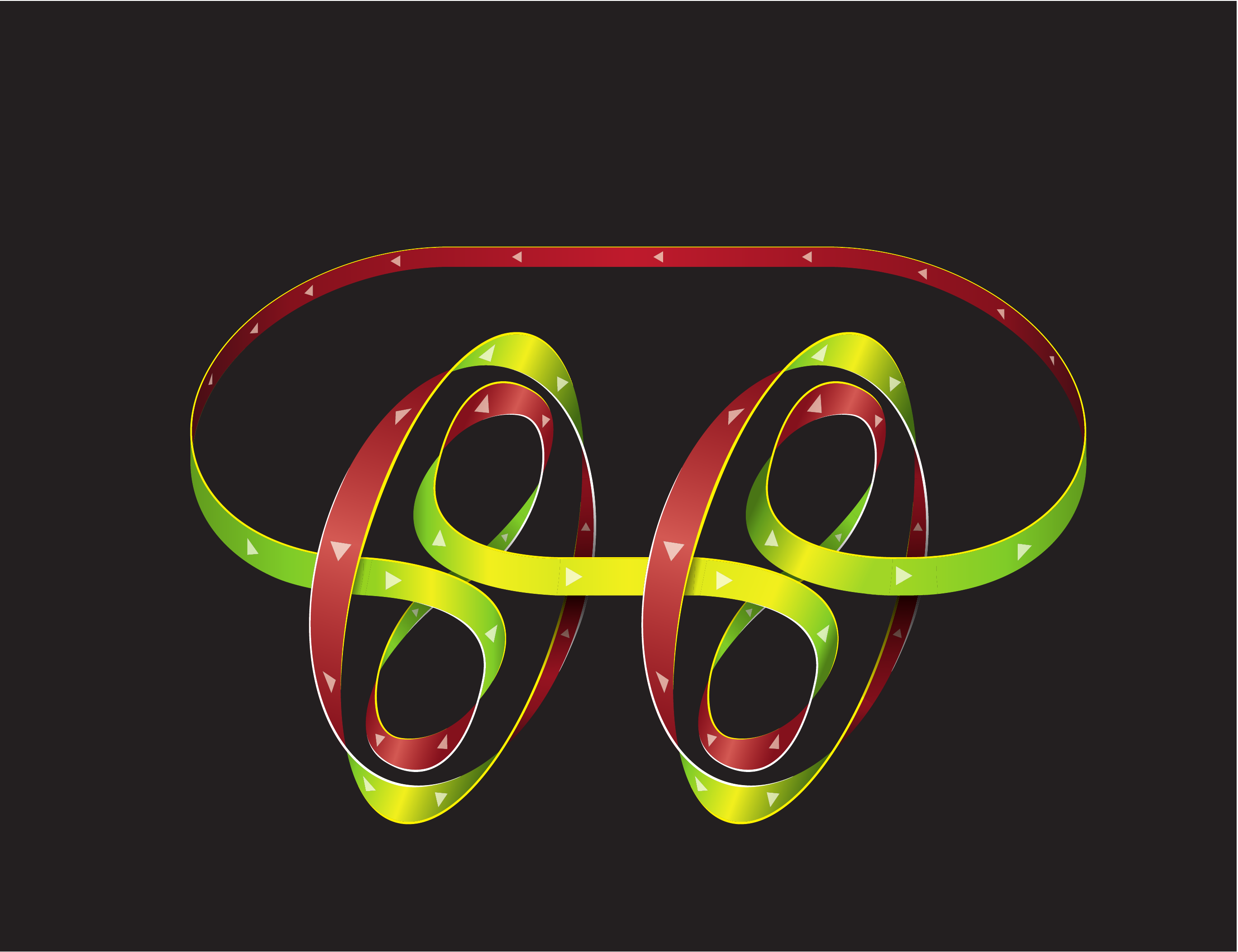}
\endminipage
\caption{Modification of Fig. 7 to include magnetic helicity conservation (adapted from Blackman and Hubbard 2014). The top left panel shows a large untwisted  toroidal magnetic field ribbon.
The top right panel shows the action of  kinetic helicity on the initial ribbon. The small scale negative kinetic helicity produces each  the four small scale poloidal loops.  Each small loop 
 incurs a writhe (or overlap) of positive (=right-handed) magnetic helicity. Since magnetic helicity is conserved, each of these four loops also has a negative (=left-handed) twist along the field ribbon.   The two intermediate scale poloidal loops  encircling the small scale loops in the bottom panel represent the resultant mean poloidal field 
averaged   separately over each pair of  loops.  
The intermediate scale poloidal loops have  accumulated two units of magnetic twist, one from each of the small loops that they encircle. These intermediate scale  loops are also  linked to the initial large scale torioidal loop.  Since a single linked pair of ribbons has  2 units of  magnetic helicity we see that the two poloidal loops linking the torioidal loop have a total 4 total units of right handed magnetic helicity in linkage which exactly balances the sum of 2+2 left handed units of twist on these poloidial loops.  In general, the small scale of the twists need not correspond to the same scale as the velocity driving the small scale writhes, though the calculations of section 4 assume such for simplicity.}
  \label{bernfig7}
\end{figure}


\begin{figure}
\minipage{.5\textwidth}
  \includegraphics[width=\linewidth]{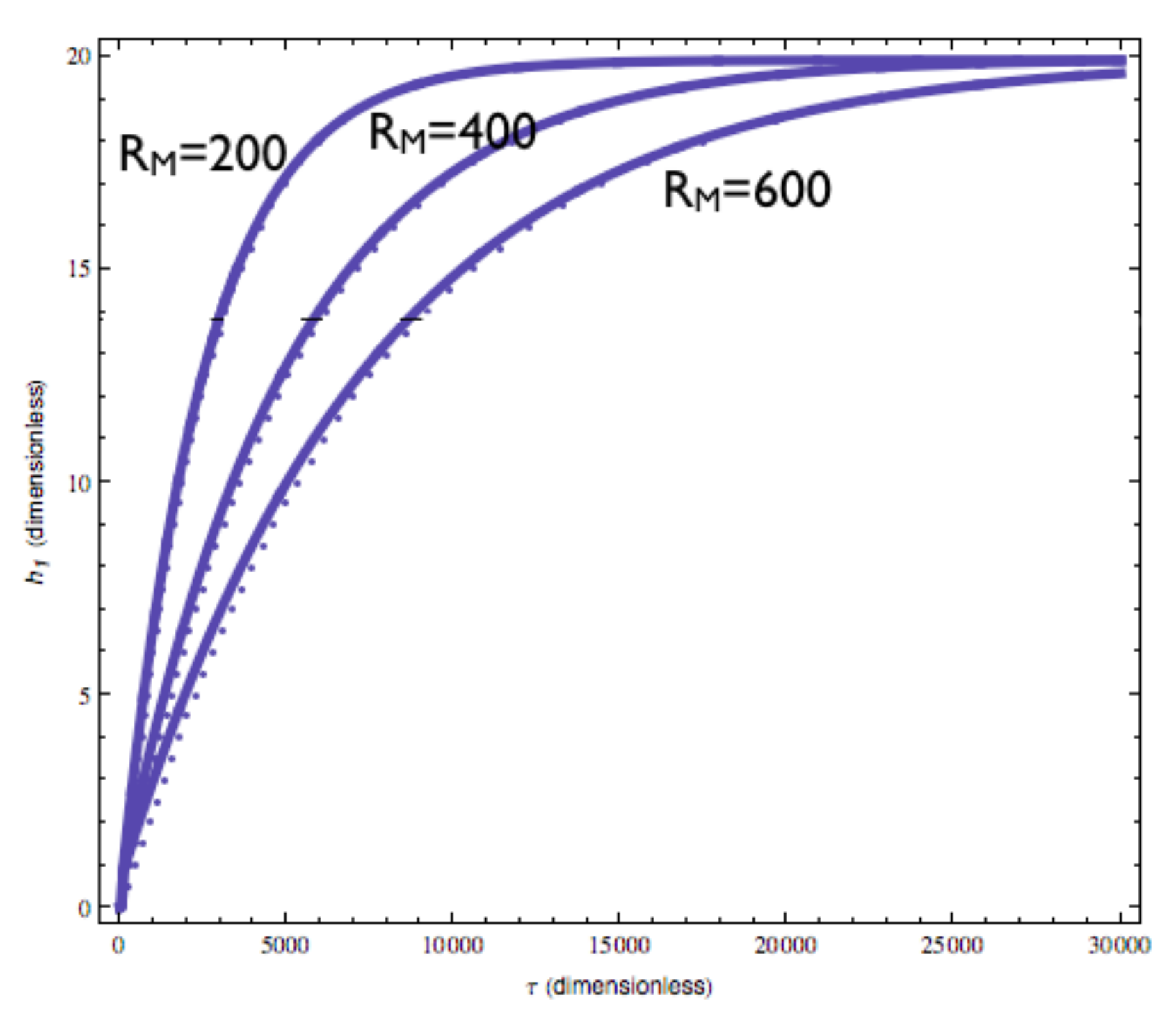}
\endminipage\hfill
\minipage{.5\textwidth}
  \includegraphics[width=\linewidth]{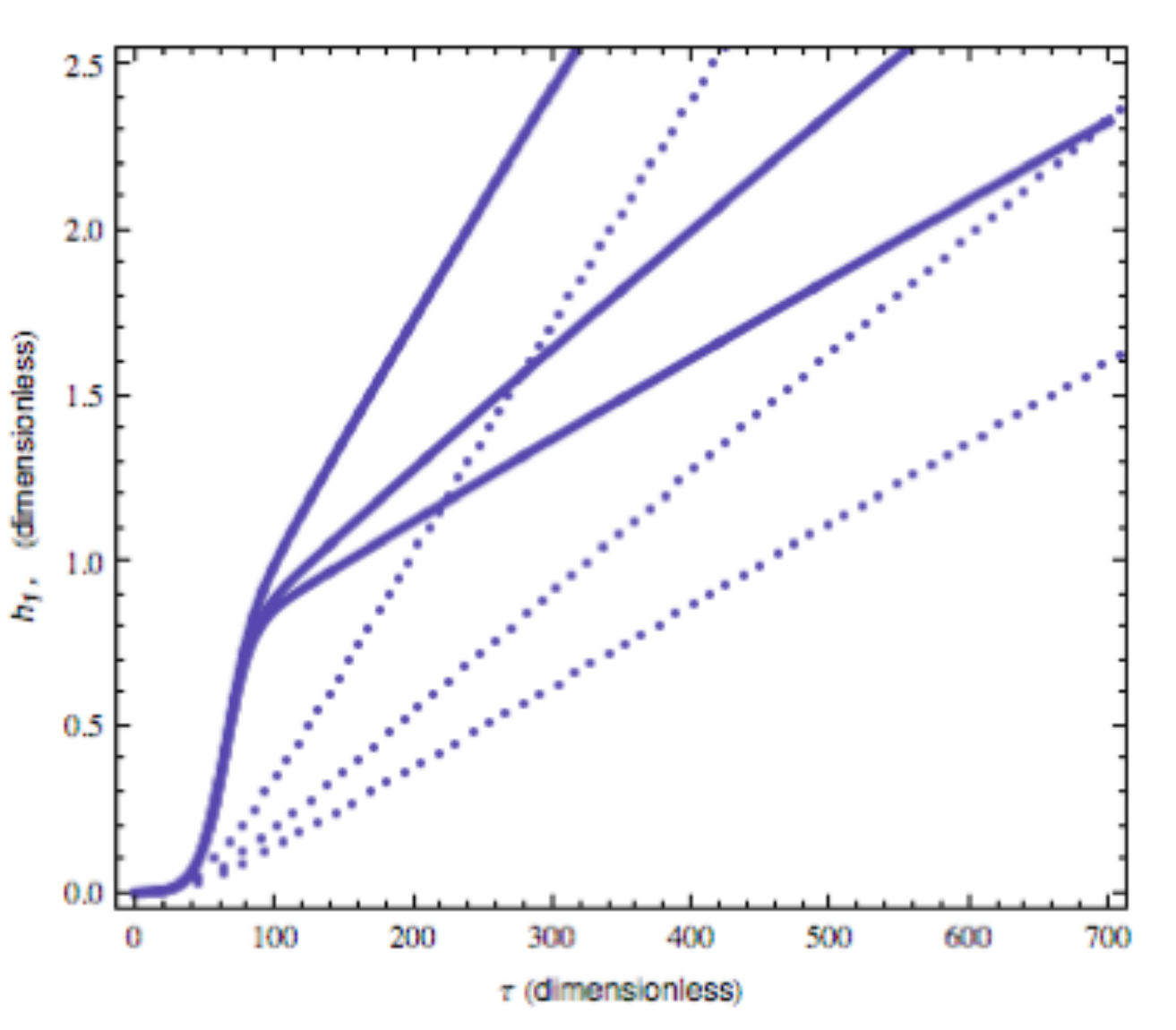}
\endminipage\hfill
\label{bernfig8}
\caption{Solutions of Eqs. (9) and (10)  for the  $\alpha^2$ dynamo problem discussed in the text 
(Figs. based on Field \& Blackman 2002 \& Blackman \& Field 2002).
   The calculation maintains $h_v$=-1  and the initial conditions are  $h_2(0)=0$ and $h_1(0)= 0.001$ and for three different magnetic  Reynolds numbers as shown (order of $R_M$ in the solid curves are the same for the right curves as the left). Time ($x$-axis) is in units of eddy turnover times at the forcing scale $k=5$.
 (a) left panel shows the solution over  a long time period highlighting the analytical result that  all curves eventually converge  toward the same final  value of $h_1$, but the higher $R_M$ cases (those which are slower to dissipate the offending small scale magnetic helicity)  take longer to get there. The dotted curves show the quasi-empirical fit formula used by 
    Brandenburg (2001) to fit simulations at late times. The dynamical theory of Eqs. (9) and (10) do very well to match this empirical
   fit formula which in turn matched  simulations.
  (b) Right panel shows the solution only through  $\tau=700$.   This highlights that at early times, before $h_2$ has grown 
  significantly, the growth of $h_1$ is independent of $R_M$.  The dotted lines are the \textit{artificial} extension of the empirical fit formula
  of Brandenburg (2001) beyond its region of validity. The   $R_M$ dependence of the dynamical solution only arises at late
  times, and the empirical fit formula is only applicable in the $R_M$ dependent regime.}
\end{figure}

\begin{figure}
\minipage{.5\textwidth}
  \includegraphics[width=\linewidth]{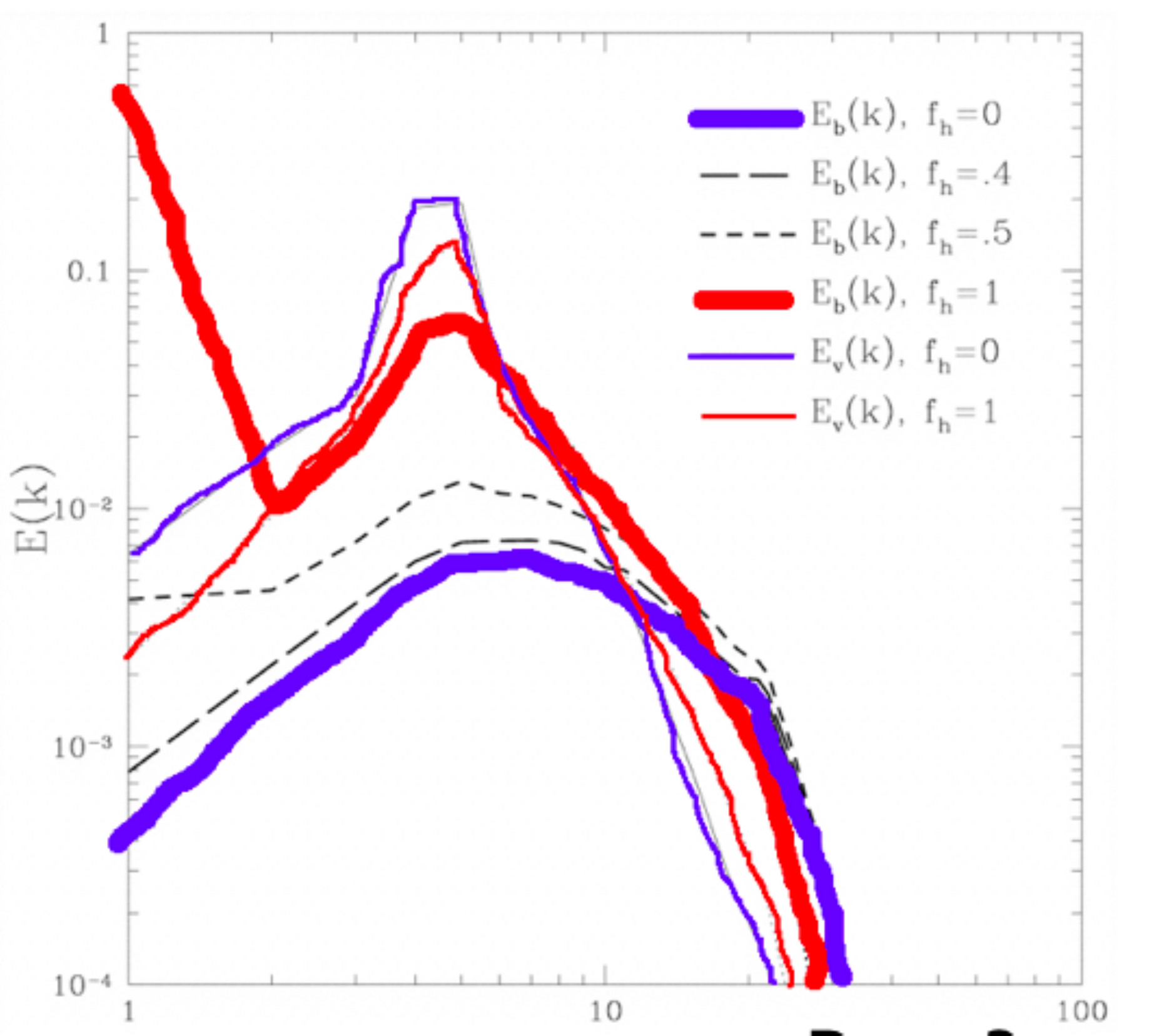}
\endminipage\hfill
\minipage{.5\textwidth}
  \includegraphics[width=\linewidth]{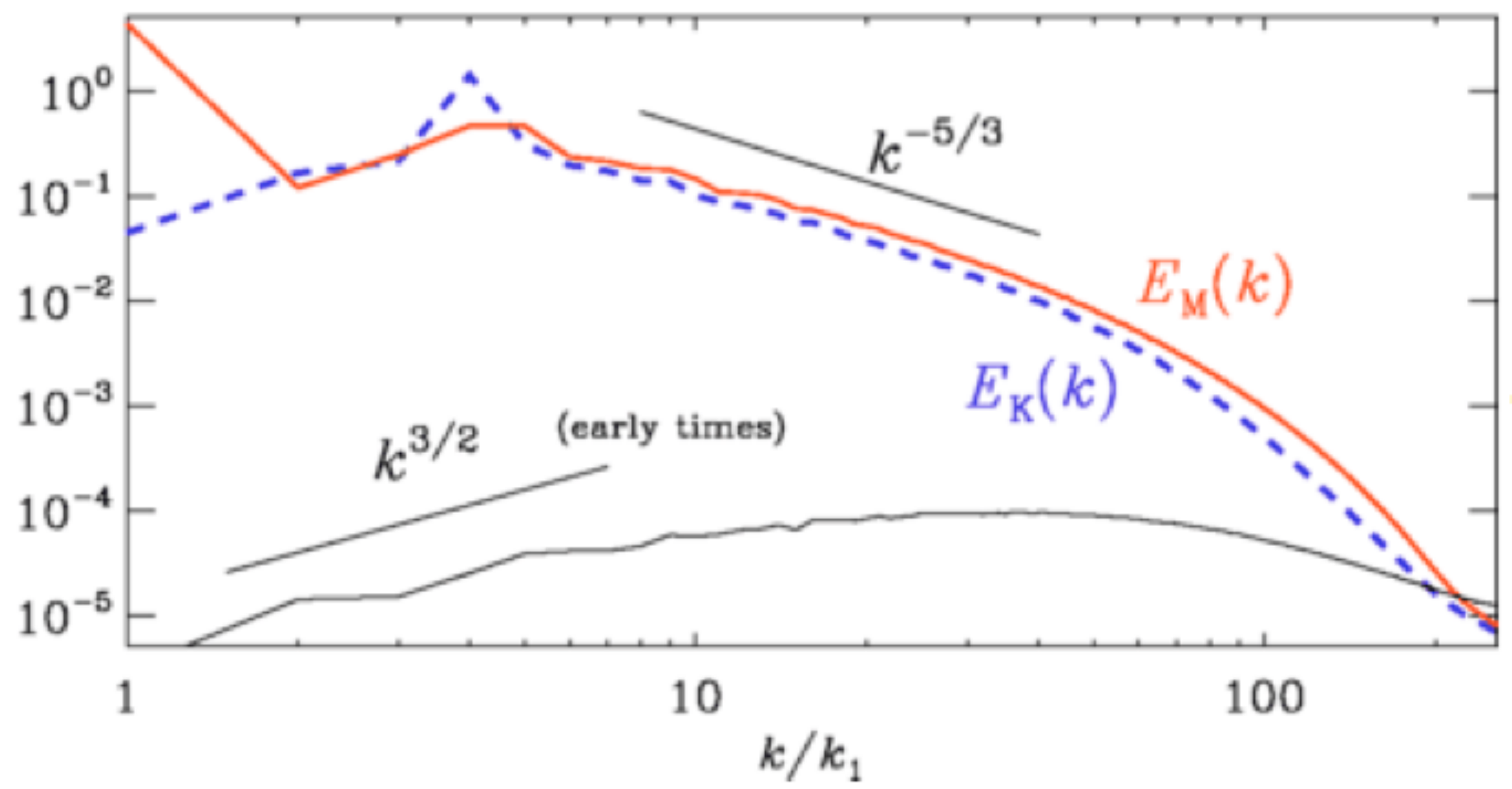}
\endminipage\hfill
\label{bernfig9}
\caption{Example steady-state saturated spectra from direct numerical simulations of helically  forced MHD turbulence.
(a) the left panel is adapted from Maron \& Blackman (2002) for $64^3$ simulation and magnetic Prandtl number 3 with forcing wavenumber
$k=5$. The thick red and blue lines are the magnetic and kinetic energy spectra  for fractional kinetic helicity $f_h=1$. The thin
 red and blue lines are the magnetic and kinetic energy spectra  for $f_h=0$. Large scale field growth (at $k=1$) is 
 dramatic in the $f_h=1$ case and negligible for $f_h=0$.
(b) The right panel is a $512^3$ simulation for $f_h=1$ for unit magnetic Prandtl number
and forcing wavenumber $k=4$  from  Brandenburg et al. (2012).   
 In the right panel, blue indicates kinetic energy  and red indicates magnetic energy.
The thick and thin red lines of the left panel thus correspond respectively to the red and blue lines of the right panel in that 
these are all for the case of $f_h=1$.    The right panel also shows  large scale $k=1$ field for helical forcing.
The essential features of the growth of the large scale field to saturation in such simulations  are captured by Eqs. (\ref{9}) and (\ref{10})
the solutions of which are shown in Fig. 9. }
\end{figure}
\newpage

\begin{figure}
\minipage{.5\textwidth}
  \includegraphics[width=\linewidth]{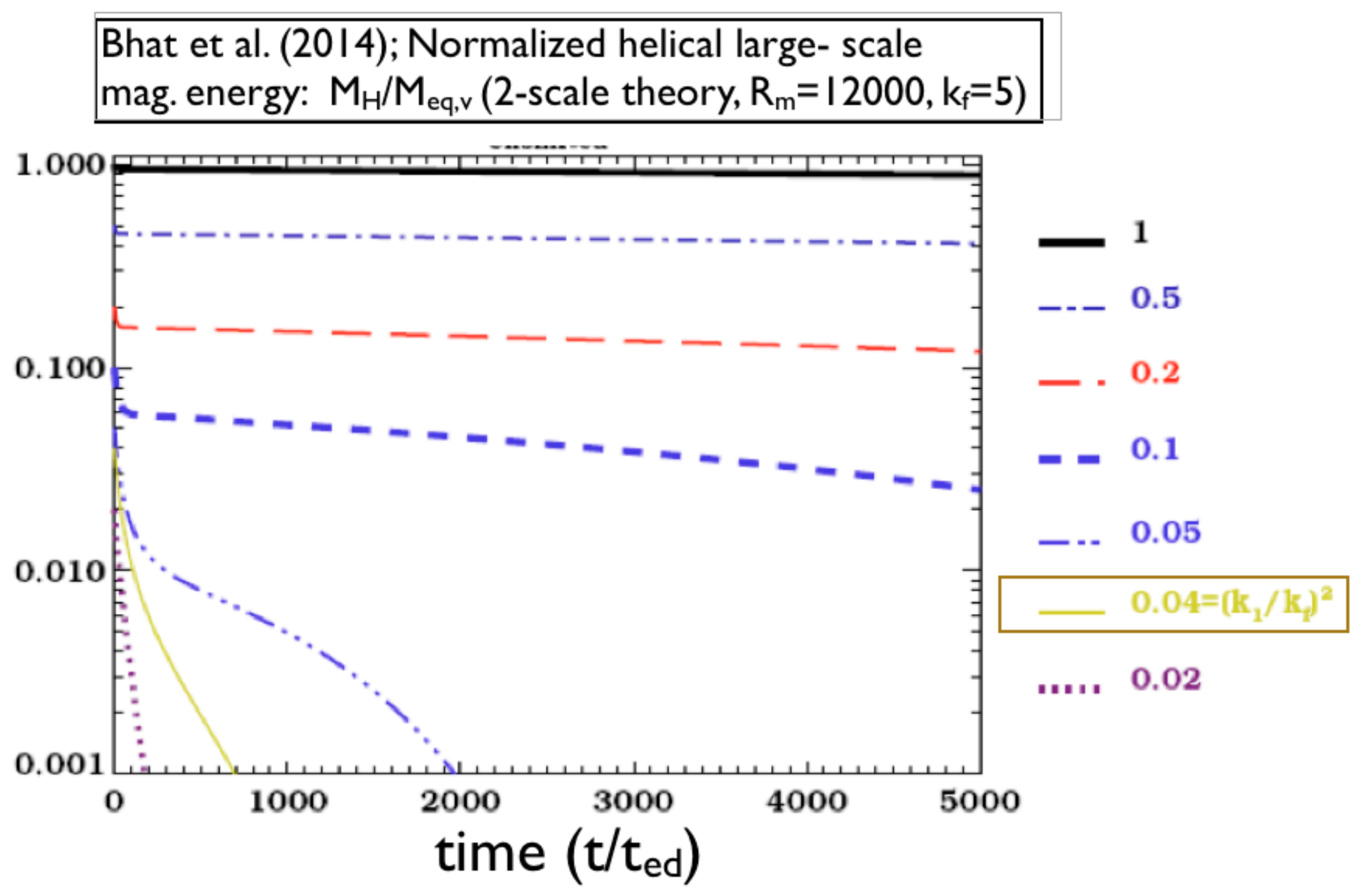}
\endminipage\hfill
\minipage{.5\textwidth}
  \includegraphics[width=\linewidth]{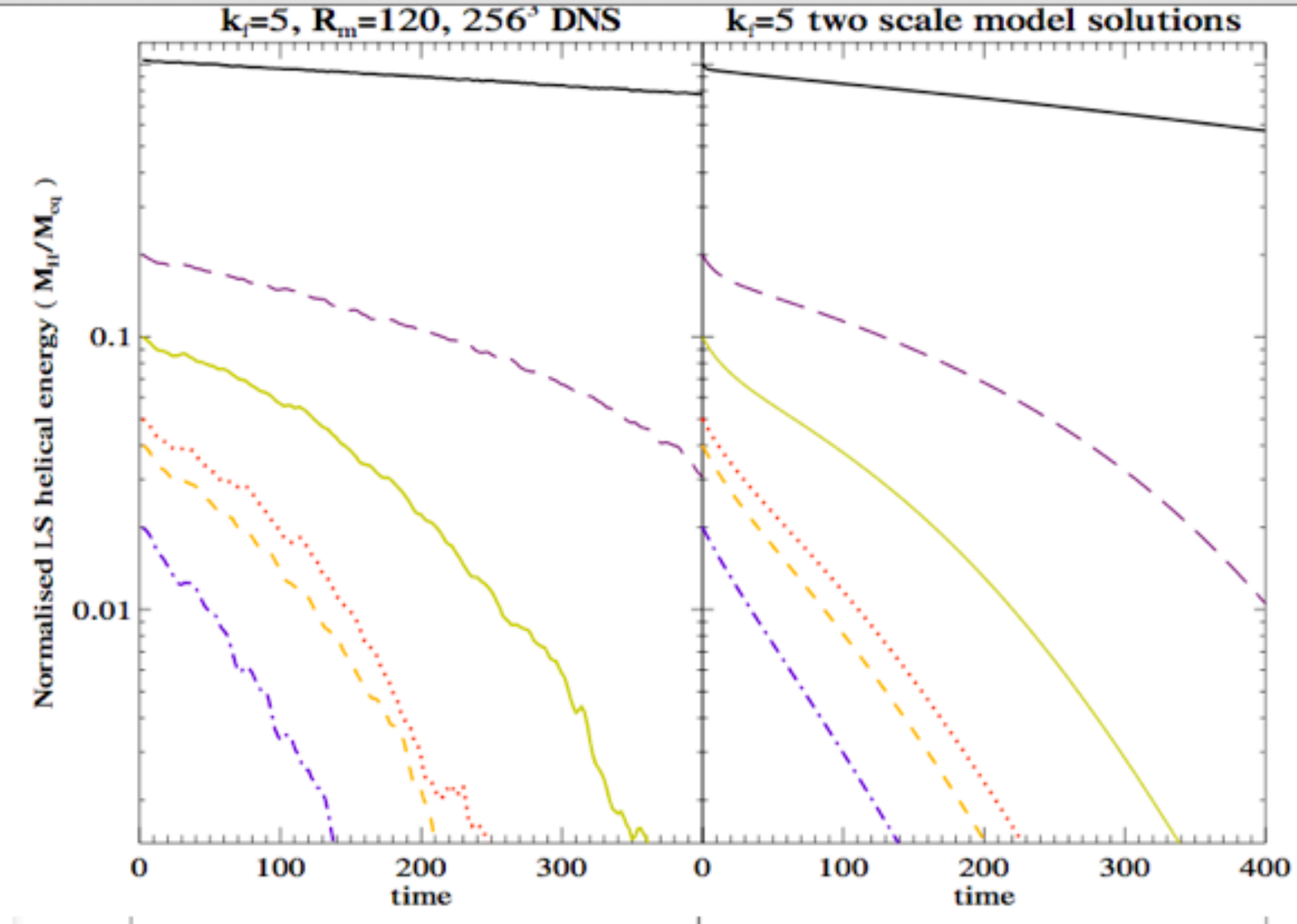}
\endminipage\hfill
\label{bernfig10}
\caption{Figure is from Bhat et al. (2014) and addresses the resilience of large scale helical fields to decay.  The left panel shows solutions of Eqs. (\ref{9}) and (\ref{10}) for  $M=k_1h_1/k_2$ ,non-dimentionalized to the turbulent equipartiton value and for  the initial values shown, subjected to steady turbulent forcing with $h_v=0$. Slow decay occurs when  initially $h_1 > k_1/k_2$ and fast decay occurs when the initial value is below this value. The critical value of $M$ is boxed in yellow. 
In the left panel, $k_2/k_1=5$ and $R_M=12000$.  The right panel shows a comparison between simulations and theory for the same problem at lower $R_M$. The agreement looks good. A subtlety however, is that the $R_M$ accessible in the simulations is too low to identify the transition value $h_1=k_1/k_2$. (see text of sec 4.3).}
\end{figure}
\newpage

\begin{figure}
\minipage{.5\textwidth}
  \includegraphics[width=\linewidth]{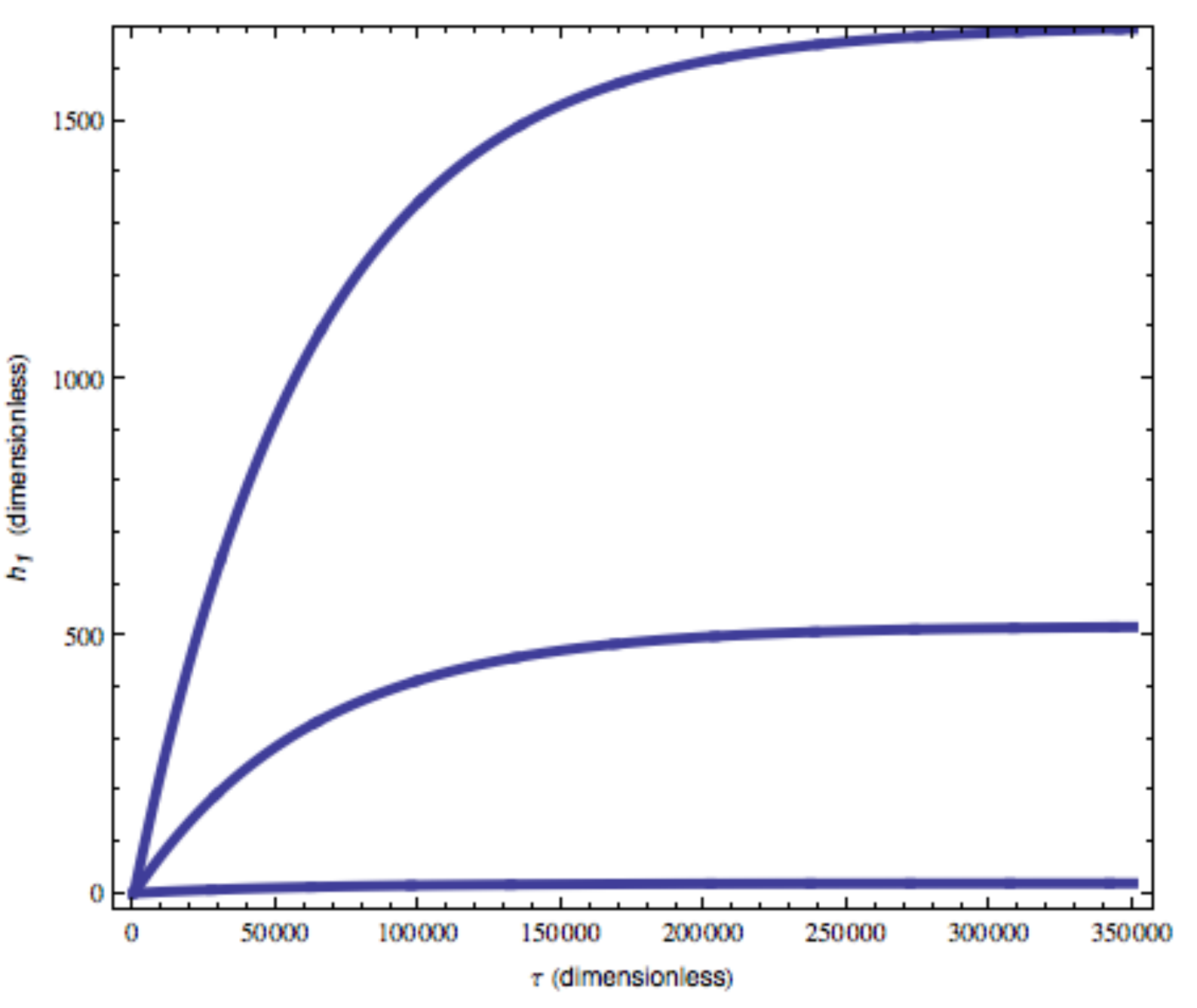}
\endminipage\hfill
\minipage{.5\textwidth}
  \includegraphics[width=\linewidth]{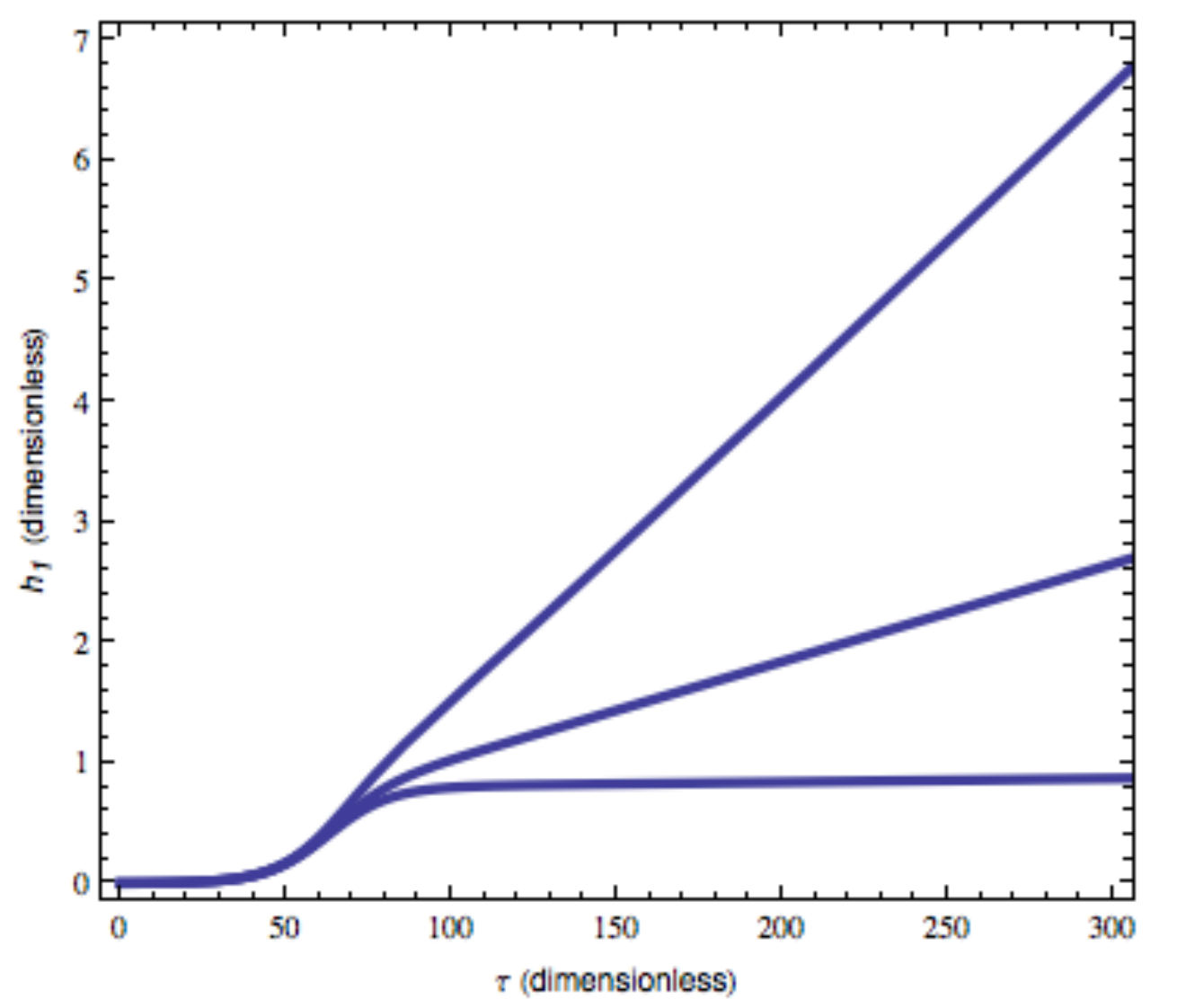}
\endminipage\hfill
  \caption{Figure updated from Blackman (2003): This shows the conceptual role of a simple advective type helicity flux in 
  increasing the $h_1$ saturation value of the $\alpha^2$ dynamo, and the trend toward extending the growth of $h_1$ before
  resistive quenching incurs.  The three curves in each panel correspond solutions of Eqs. (\ref{9}) and  (\ref{10}) modified by the  addition of a loss term  $- \lambda h_2$ to Eq. (\ref{10}). All curves correspond to $R_M=5000$. From top to bottom  in each panel the curves have $\lambda=1/30, 1/100$ and $0$ respectively. Left panel (a) is the late time regime: The bottom curve saturates at the same value as those in Fig. 8a, as  those have no loss term (although Fig. 8a has faster growth because the $R_M$ values are smaller). The right panel (b) is the early time regime, and shows that for $\lambda=1/30$  the resistive turnover in the $h_1$ curve is nearly avoided. This holds true even more dramatically for all larger values of $\lambda$ (not shown).}
    \label{bernfig11a}
\end{figure}

\begin{figure}
  \includegraphics[width=\columnwidth,trim=0 0 0 0]{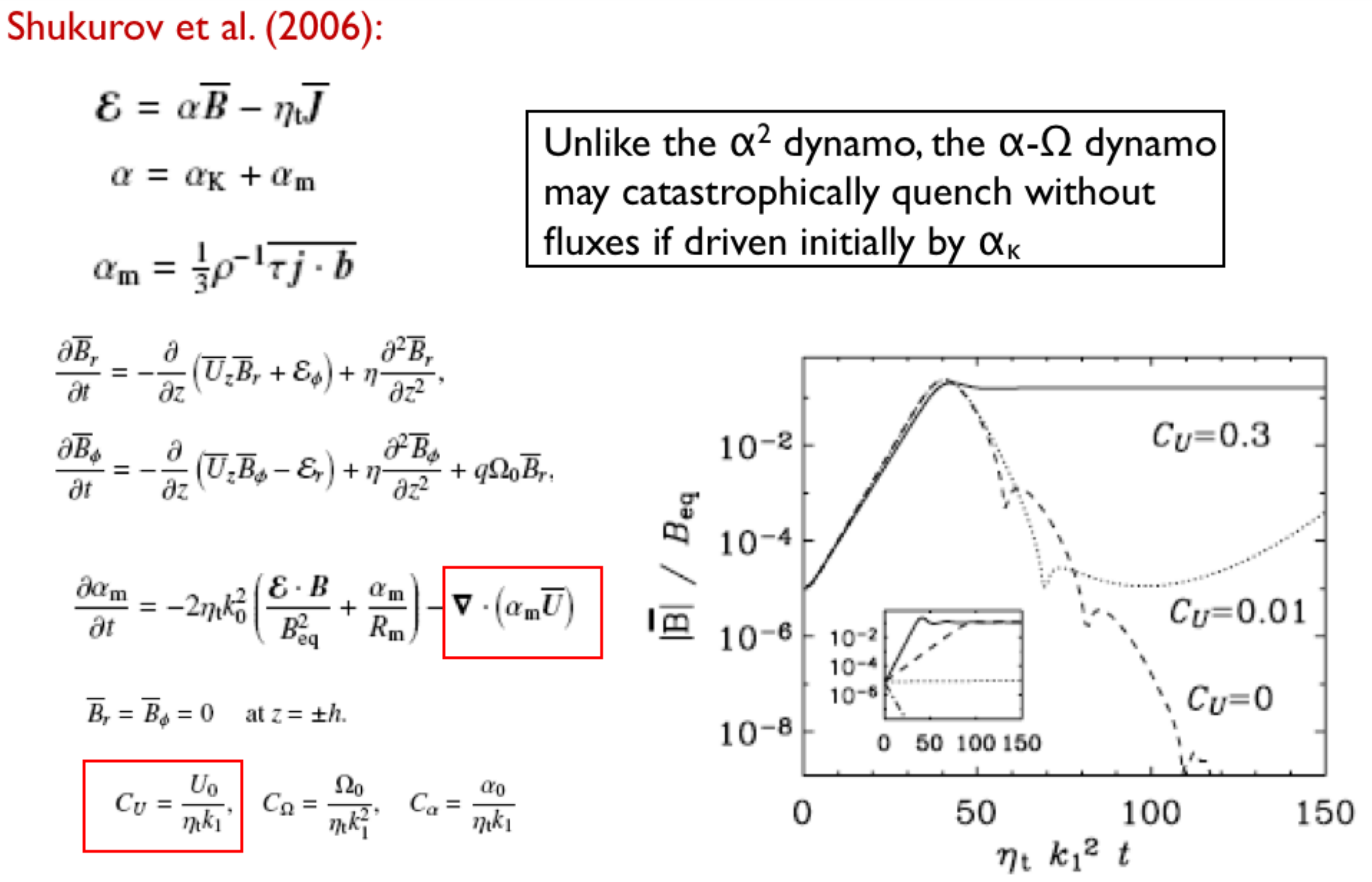}
  \caption{From Shukurov et al. (2006), this provides a simple model illustrating the importance of helicity fluxes in sustaining the $\alpha-\Omega$ dynamo when magnetic helicity dynamics and fluxes are coupled into the theory. They considered an advective contribution to the helicity flux, as shown boxed in red, and quantified its contribution by $C_U$. The plots on the right show that only above a threshold value of $C_U$ does the large scale mean field (plotted in units of the field strength corresponding to equipartition with the turbulent kinetic energy) sustain. Without these fluxes, the field decays. This contrasts the situation of the $\alpha^2$ dynamo in Figure 11 in which flux terms increase the saturation value but do not determine the difference between growth and decay.}
    \label{bernfig11}
\end{figure}

\begin{figure}
  \includegraphics[width=\columnwidth,trim=0 0 0 0]{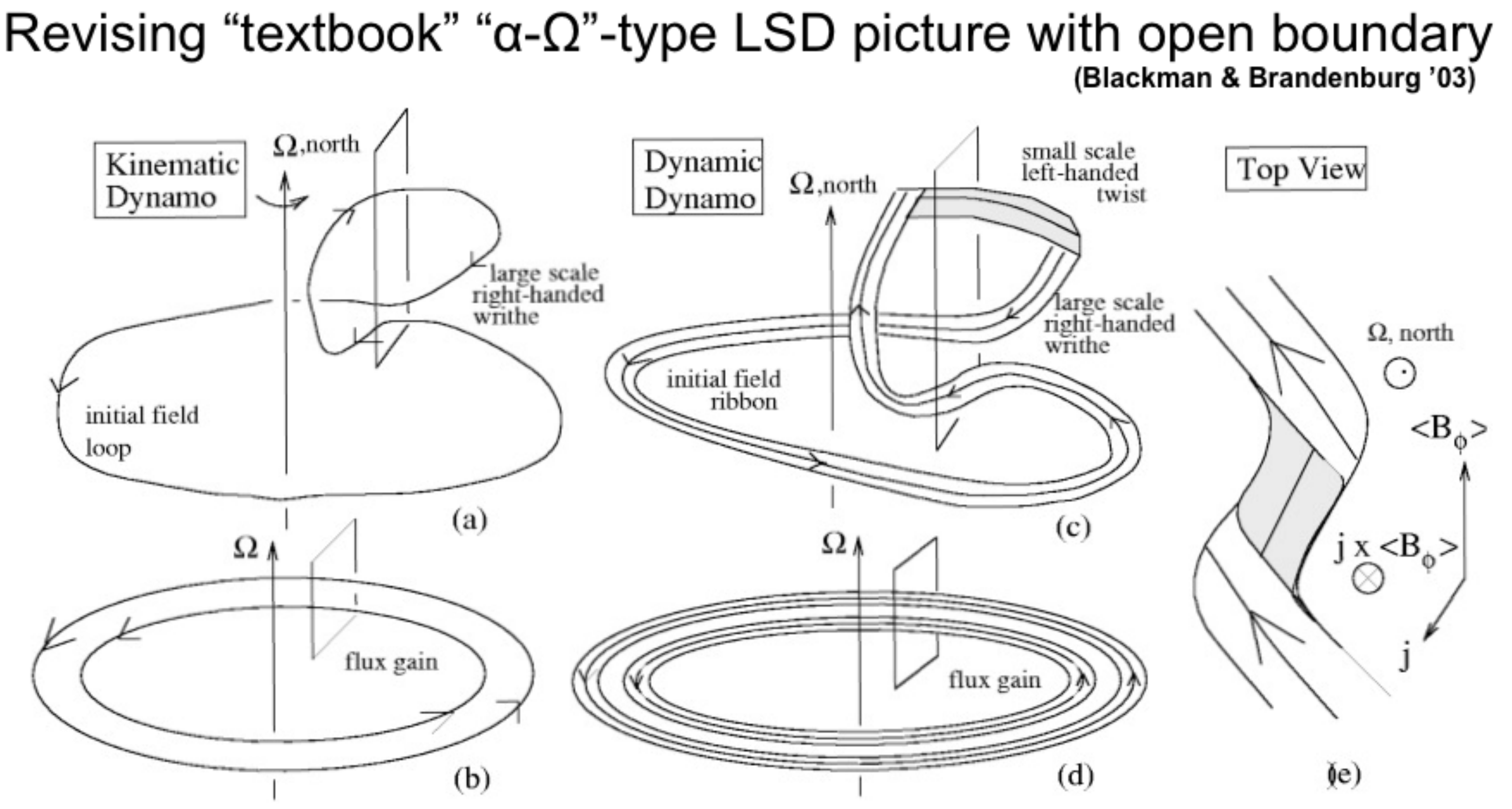}
  \caption{From Blackman \& Brandenburg (2003),  a comparison of the classic picture of $\alpha-\Omega$ dynamo without magnetic helciity conservation (panel a)  to that with magnetic helicity conservation (panel c).  
  Panels (b) and (d) show the gain in toroidal field after ejection of the large poloidal  loop. For panel  (d), this ejection alleviates the twist that would otherwise build up.
  This figure is a conceptual generalization of the concepts addressed by the comparison of  Figs 5 and 6 to include shear and buoyancy and to motivate why  e.g. coronal mass ejections of the sun, or galaxies may represent the irreversible loss of small scale magnetic helicity that allows LSD action to sustain. See also Fig. 10 }
  \label{bernfig12}
\end{figure}

\begin{figure}
  \includegraphics[width=\columnwidth,trim=0 0 0 0]{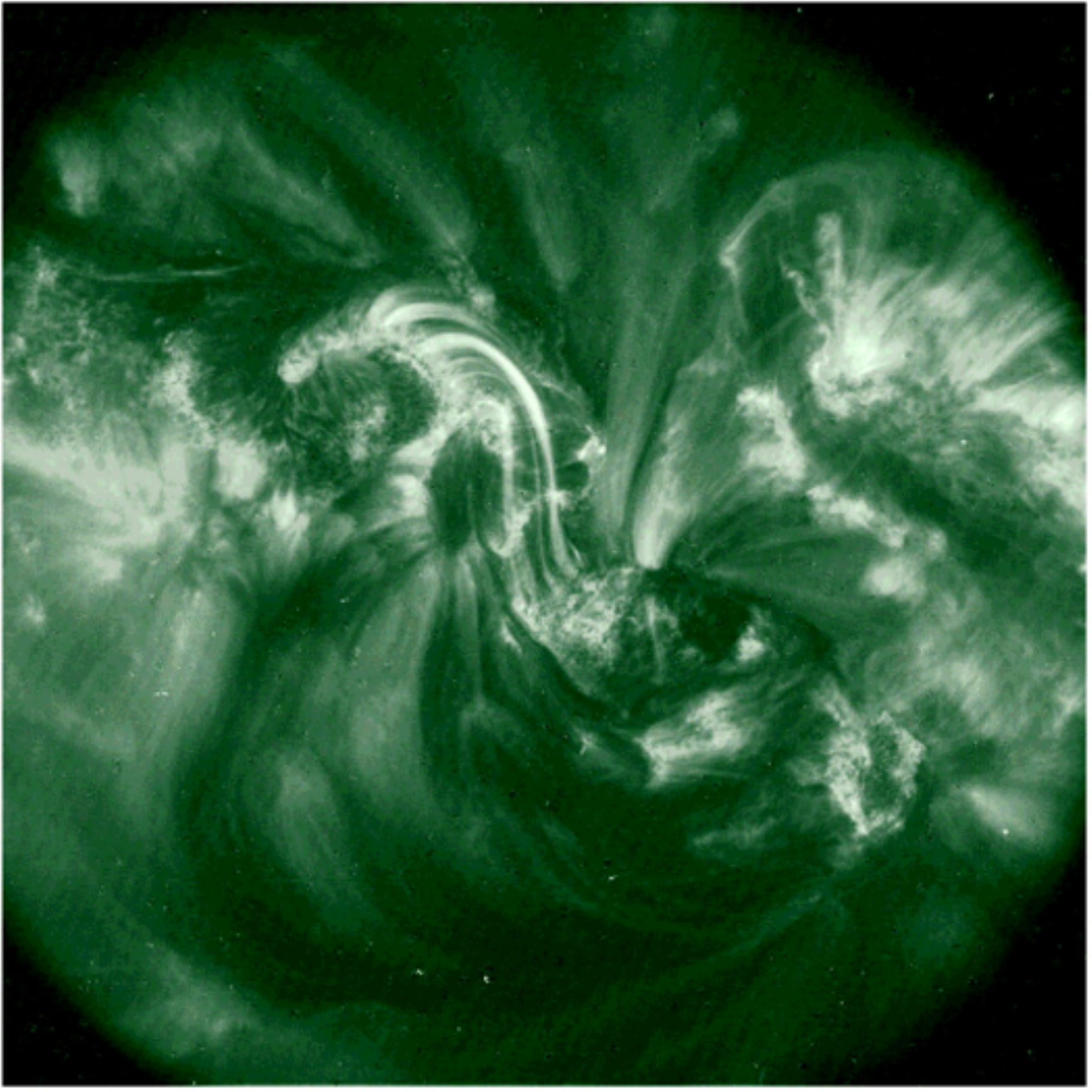}
  \caption{TRACE 195\AA image of a solar sigmoid of the Northern hemisphere from Gibson et al. (2002). It is tempting to interpret the large scale right-handed writhe and small scale left handed striations along the sigmoid being consistent with what a magnetic helcity conserving dynamo would predict when the fluxes eject both small and large scale helicities from the interior. The ejection of the small scale helicity in sigmoids or coronal mass ejections may be fundamental to the operation of the solar dynamo, not just a consequence of large scale field generation.}
  \label{bernfig13}
\end{figure}

\begin{figure}
  \includegraphics[width=\columnwidth,trim=0 0 0 0]{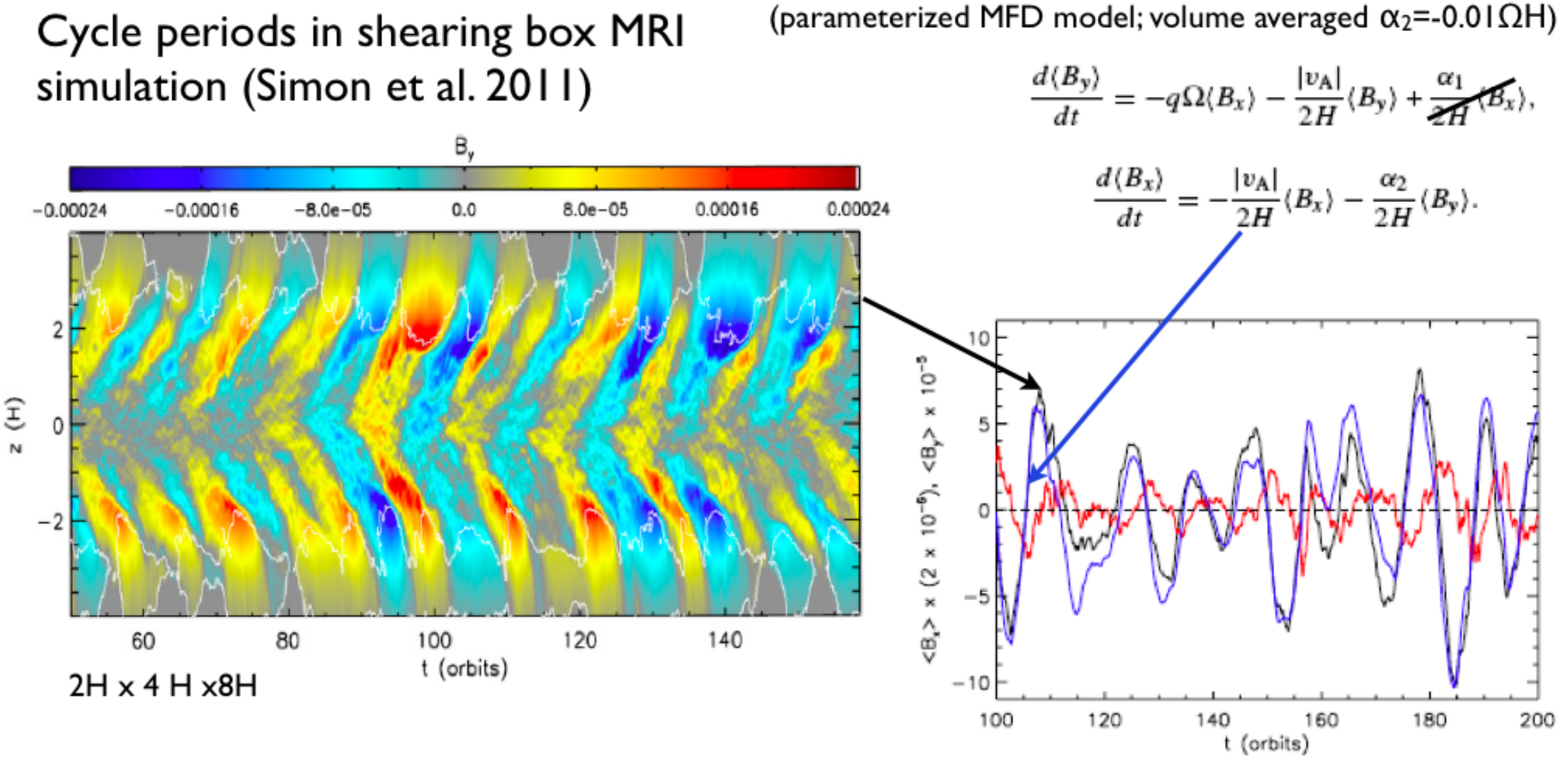}
  \caption{Example of evidence for large scale dynamo action in  MRI simulations from Simon et al. (2011), compared with an empirically tuned $\alpha-\Omega$ dynamo model. The left panel shows the net torioidal field vs. time,  volume  averaged over all $x$ and $y$ and  over $|z|<0.5H$. Outflow vertical boundaries were used and periodic boundaries in the other two dimensions. The cycle period of $\sim 10$ orbits is evident. The black line in the  right panel shows the mean  toroidal field  as function of time and the blue line corresponds to the model semi-empirical fit equation of the $\alpha-\Omega$ dynamo.  The sign of the required dynamo $\alpha$ coefficient is opposite to that expected from kinetic helicity.  }
  \label{bernfig14}
\end{figure}

\begin{figure}
  \includegraphics[width=\columnwidth,trim=0 0 0 0]{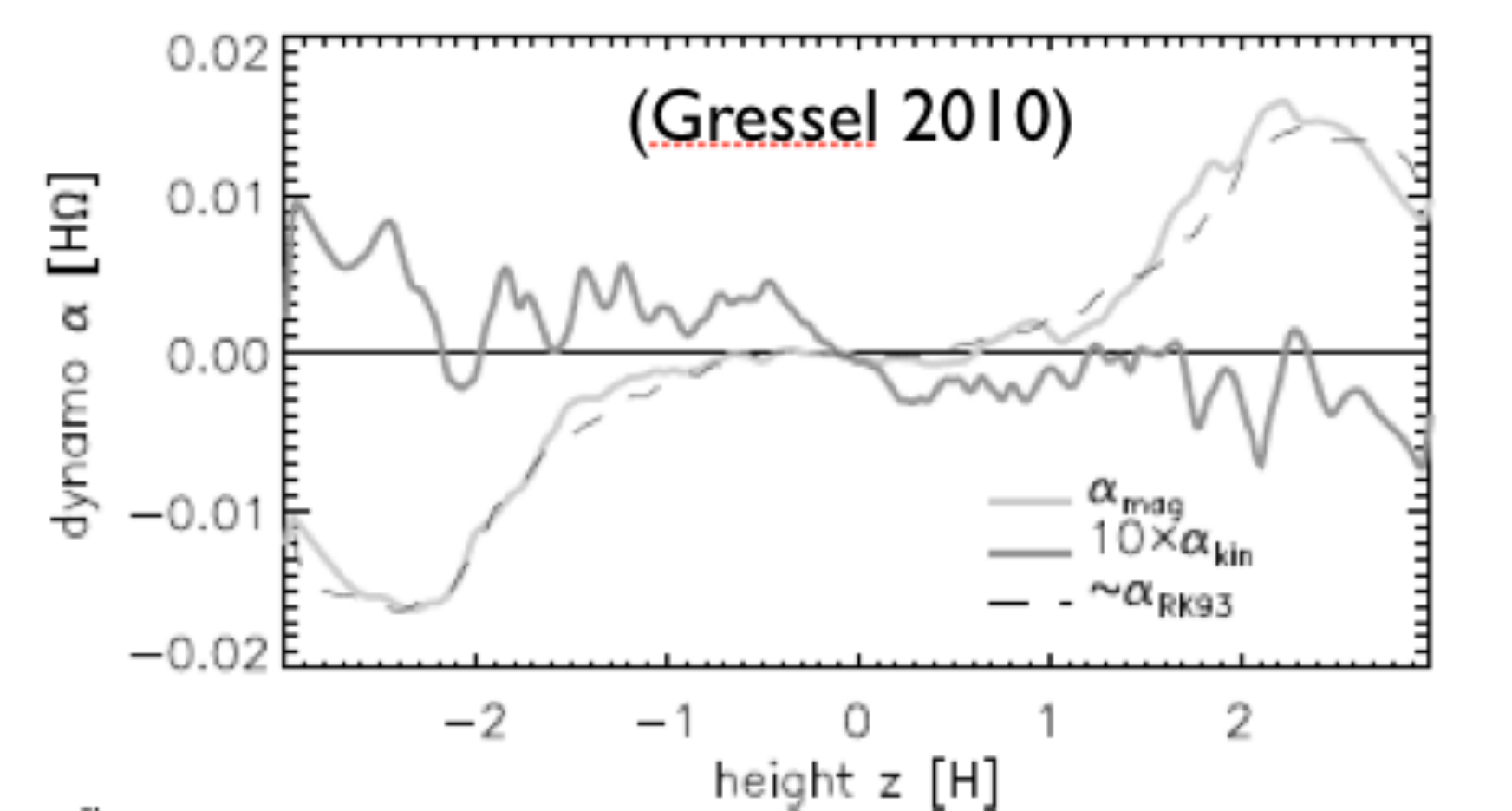}
  \caption{Example of $\alpha-\Omega$ dynamo model for large scale fields compared to vertically stratified  shearing box, MRI simulations from Gressel (2010). The vertical dependences of the $\alpha$ dynamo coefficients are  shown. This coefficient is the proportionality between
  the EMF and the mean field (averaged in radius and azimuth). The values $\alpha_{kin}$ and $\alpha_{mag}$ are proportional to the kinetic and magnetic helciities respectively. The value $\alpha_{RK93}$ comes from R{\"u}diger \& Kitchatinov (1993) and is derived for stratified rotating turbulence. The figure shows that  $\alpha_{RK93}$ or  $\alpha_{mag}$ are much better fits to the dynamo in the shearing box than the traditional $\alpha_{kin}$ of 20th century textbooks.  It would seem that $\alpha_{RK93}$ is therefore capturing
  the $\alpha_{mag}$ contribution and may result from magnetic buoyancy.}
  \label{bernfig15}
\end{figure}

\begin{figure}
  \includegraphics[width=\columnwidth,trim=0 0 0 0]{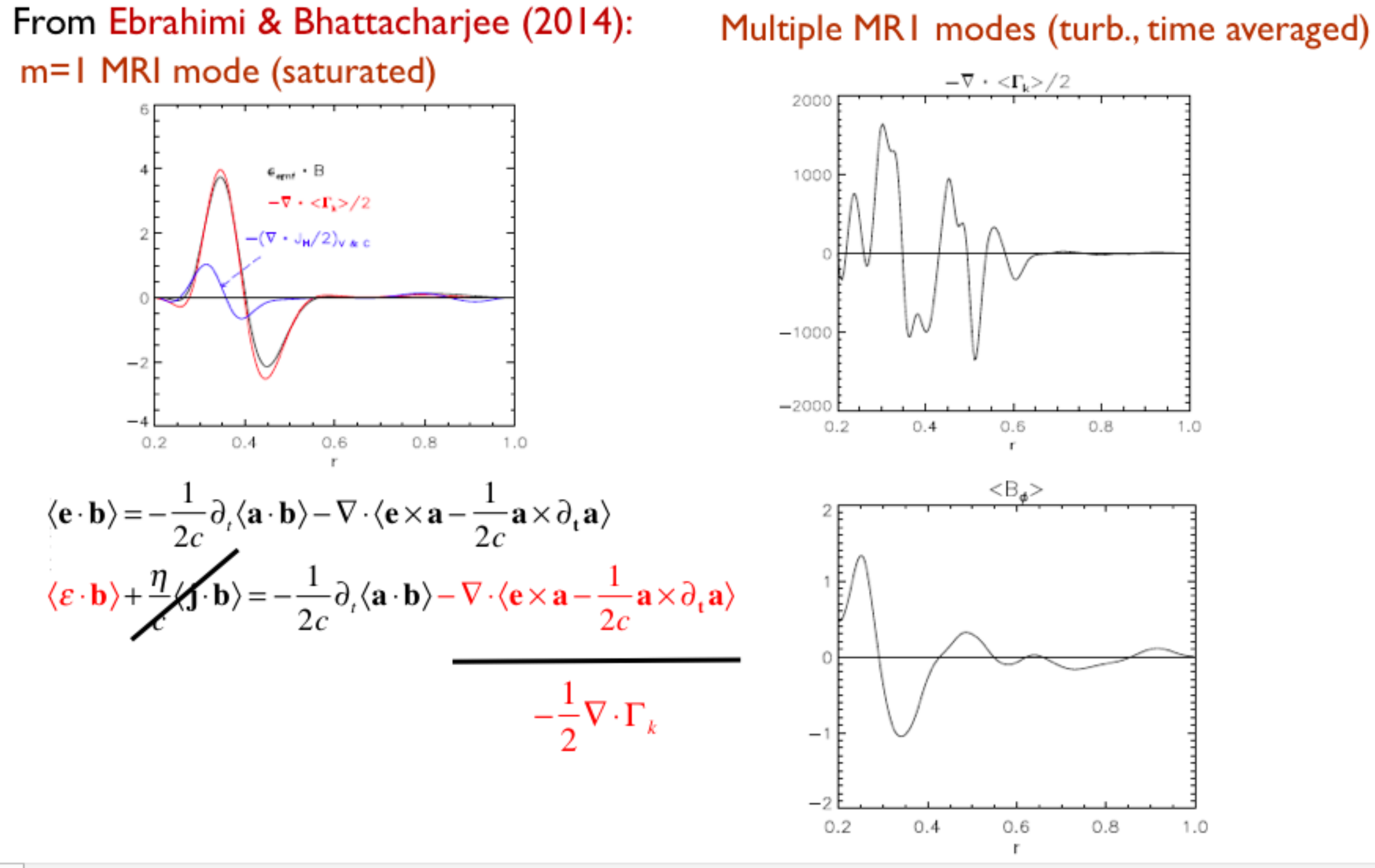}
  \caption{Figures  adapted  from Ebrahimi \& Bhattacharjee (2014), representing    $R_M\gg1$,  and  unit magnetic Prandtl number,  
   shear flow driven simulations of an MRI unstable cylinder with perfectly conducting boundaries. This  shows evidence for the generation of large scale field when averages are taken over height and azimuth, leaving the radial variable unaveraged.  No initial net torioidal field was present in the box but one  emerges as a result of LSD action. 
  (a) Left panel: strong evidence for the importance of the radial flux of small scale magnetic helicity in sustaining  the EMF needed for dynamo action 
  for a single saturated unstable MRI  mode  is shown. The black curve is a measure of the field aligned EMF and the red curve is the helicity flux in shown in the equation. The blue curve is the Vishniac-Cho (2001) flux which is too small to match the total EMF sustaining  flux.  (b) Right pair of figures shows the radial correspondence between the net toroidal field and helicity flux divergence additionally time-averaged for a turbulent state in which multiple MRI unstable modes interact. The correspondence  provides further evidence for the importance of local helciity fluxes in sustaining the LSD.}
  \label{bernfig16}
\end{figure}

\end{document}